\documentclass[11pt,letterpaper]{article}
\pdfoutput = 1
\usepackage     {jheppub}
\usepackage     [mathscr]{eucal}
\usepackage     {bm}
\bibpunct	{[}{]}{,}{n}{}{;}

\def\tocsqueeze	{\vspace*{-3.7pt}}

\title          {    Universal hydrodynamic flow in holographic planar shock collisions }
\author[a]      {Paul~M.~Chesler,}
\author[b]	{Niki~Kilbertus,}
\author[c]	{Wilke~van~der~Schee}
\affiliation[a]	{Department of Physics, Harvard University, Cambridge MA 02138, USA}
\affiliation[b]	{Institut f\"ur Theoretische Physik, Universit\"at Regensburg, D-93040 Regensburg, Germany}
\affiliation[c] {Center for Theoretical Physics, MIT,
                Cambridge MA 02139, USA}
\emailAdd	{pchesler@physics.harvard.edu}
\emailAdd{niki.kilbertus@physik.uni-regensburg.de}
\emailAdd       {wilke@mit.edu}

\abstract
    {%
    We study the collision of planar shock waves in AdS$_5$ as a function of shock profile.
    In the dual field theory the shock waves describe planar sheets of energy
    whose collision results in the formation of a 
    plasma which behaves 
    hydrodynamically at late times.  
    We find that the post-collision stress tensor near the light cone exhibits transient non-universal behavior which depends on both
    the shock width and the precise functional form of the shock profile.
    However, 
    over a large range of shock widths, including 
    those which yield qualitative different behavior near the future light cone,
    and for different shock profiles, we find universal 
    behavior in the subsequent hydrodynamic evolution.  
    Additionally, we compute the rapidity distribution of produced 
    particles and find it to be well described by a Gaussian.
    }

\begin{document}

\advance\textheight 55pt
\maketitle
\thispagestyle{empty}
\advance\textheight -55pt

\addtocontents	{toc}{\tocsqueeze}
\addtocontents	{toc}{\tocsqueeze}
\section	{Introduction and summary}

Holographic duality \cite{Maldacena:1997re,Witten:1998qj,Gubser:1998bc} has proven to be a useful tool to study the dynamics 
of strongly coupled quark-gluon plasma (for a review see for example \cite{CasalderreySolana:2011us}).  One interesting
problem is the collision of gravitational shock waves in AdS$_5$, which can result in the formation of a black hole.
In the dual field theory, which is $\mathcal N = 4$ supersymmetric Yang-Mills (SYM), colliding gravitational waves are equivalent to colliding distributions of energy,
which for brevity we simply refer to as shock waves.
The formation of a black hole is dual to the formation of a quark-gluon plasma and the ring down of the black hole encodes the relaxation of the plasma
to a hydrodynamic description. 
The complete evolution of the field theory stress tensor $T^{\mu \nu}$ --- from pre-collision dynamics to far-from-equilibrium dynamics to hydrodynamics --- 
is encoded in the dual classical relativity problem.  

Unlike QCD, where nuclei are bound states whose energy distribution is fixed by the theory, in conformal SYM the energy distribution of colliding shocks
is not fixed;  shock energy distributions can have any desired shape and only must propagate at the speed of light.  
In the dual gravitational description this reflects the
fact that gravitational waveforms are not fixed by Einstein's equations and gravitational waves always propagate at the speed of light.
This has led to studies of a diverse range of energy profiles from planar shocks with $\delta$-function longitudinal profiles 
\cite{Grumiller:2008va,Albacete:2008vs,Albacete:2009ji} to planar shocks with finite longitudinal thickness \cite{Chesler:2010bi, Casalderrey-Solana:2013aba,Casalderrey-Solana:2013sxa,Chesler:2013lia} to shocks which are also localized in the plane transverse to the collision axis \cite{Gubser:2008pc,Chesler:2015wra,Gubser:2009sx,Lin:2009pn}
to holographic models of proton-nucleus collisions \cite{Chesler:2015bba}.

The fact that SYM doesn't specify the energy distribution of colliding shocks begs the question: 
how much do details in different energy distributions imprint themselves on the 
future hydrodynamic evolution of the produced quark-gluon plasma?
What features of collisions in SYM are universal and what features depend on one's 
chosen energy profile for the shocks?  Indeed, in \cite{Casalderrey-Solana:2013aba},
where the collision of planar shocks with Gaussian longitudinal profiles was studied, it was found that 
some qualitative features of the debris produced by the collision are sensitive to the thickness of the shocks.
For suitably thin shocks remnants of the initial shocks can survive the collision event
and propagate on the forward light cone and regions of negative energy density appear near the light cone.
In contrast, when the shock thickness is suitably large no obvious remnants of the initial shocks survive the collision 
event and the energy density is everywhere smooth and positive in the forward light cone \cite{Chesler:2010bi,Casalderrey-Solana:2013aba}.

To begin to address the above questions we focus on the simple case of planar shock collisions in AdS$_5$,
where the shocks have no dependence on the coordinates transverse to the collision axis.  The shocks move in the $\pm z$ direction
at the speed of light and have energy density 
\begin{equation}
\label{eq:energystress0}
T^{00} = \frac{N_{\rm c}^2}{2 \pi^2} \mu^3 \delta_w(z\pm t),
\end{equation}
with $t$ time, $N_{\rm c}$ the number of colors, and $\mu$ an energy scale.
We consider $\delta_w(x)$ which are ostensibly smeared $\delta$-functions localized about $x = 0$ with normalization and variance 
\begin{align}
\label{eq:normandvar}
&\int dx \,\delta_w(x) = 1,&
&\int dx \, x^2 \delta_w(x) = w^2.
\end{align}
Hence, the energy per unit transverse area of the shocks is $\frac{N_{\rm c}^2}{2 \pi^2} \mu^3$.  
We investigate the collision dynamics 
as a function of the shock width and the functional form of $\delta_w$.
For simplicity we consider profiles $\delta_w$ in which $w$ is the only scale.
The dimensionless measure of the shock width is  $\mu w$ for which we consider $\mu w \lesssim \frac 12$.
In contrast to the collisions studied in \cite{Chesler:2010bi,Casalderrey-Solana:2013aba}, which were
in the background of a low temperature plasma,
we study collisions at  zero background temperature.
This allows us to study long time evolution without pollution due to thermal regulators.

We find that the post-collision stress tensor near the light cone is non-universal and depends on both
the shock width $w$ and the precise functional form of the shock profile $\delta_w$.  
However, we observe that the non-universal behavior is transient:
irrespective of $w$ or $\delta_w$, long after the collision event, nearly all the energy 
lies inside the future light cone and the evolution of the stress tensor is governed by hydrodynamics.
Over a large range of shock widths, including 
those which yield qualitative different behavior near the future light cone,
and for different shock profiles $\delta_w$, we find universal 
behavior in the initial hydrodynamic data.  
On a surface of constant proper time $\tau = \tau_{\rm init} \gtrsim t_{\rm hydro}$, with $t_{\rm hydro} \approx 2/\mu$ the hydrodynamization time,  
we find that the fluid velocity 
is well described by boost invariant flow and that the proper energy density $\epsilon$ is well described by
\begin{equation}
\label{eq:gaussianform0}
\epsilon(\xi,w)|_{\tau = \tau_{\rm init}} = \mu^4 A(\mu w) f\left (\textstyle \frac{\xi}{\xi_{\rm FWHM}(\mu w)} \right ),
\end{equation}
where $\xi$ is spacetime rapidity and $f$ is a $w$-independent function with $\xi_{\rm FWHM}$ its full width at half maximum.
Therefore, the only $w$ dependence in the hydrodynamic flow is in the proper energy's normalization and rapidity width.
Aside from this $w$ dependence, we observe that the hydrodynamic flow is otherwise insensitive to the precise functional form of $\delta_w$.

Choosing $\tau_{\rm init}  = 3.5/\mu$ and the normalization condition $f(0) = 1$,
we find $A(\mu w )$ and $\xi_{\rm FWHM}(\mu w)$ are well approximated by the quadratic functions 
\begin{subequations}
\label{eq:quadfits}
\begin{eqnarray}
 A(\mu w) &=& 0.14 + 0.15 (\mu w) - 0.025( \mu w)^2,
\\
\xi_{\rm FWHM}(\mu w) &=& 2.25 - 1.15 (\mu w) + 0.31 (\mu w)^2.
\end{eqnarray}
\end{subequations}
Note $\frac{d\xi_{\rm FWHM}}{d (\mu w)} \approx -1$, signifying appreciable $w$ dependence.
The function $f$ is well described by a Gaussian with unit full width at half maximum, 
\begin{equation}
\label{eq:unitgaussian}
f(x) = e^{-\frac 12 x^2/\sigma^2}, \ \ \ \sigma = [8 \log 2]^{-1/2} \approx 0.425.
\end{equation}

Given that the $w\to 0$ limit of our collisions is that of colliding $\delta$-functions, it is not 
surprising that the hydrodynamic evolution becomes insensitive to both $w$ and the details 
of the shock profile when $w$ is sufficiently small.  Indeed, similar insensitivities were 
observed in \cite{Casalderrey-Solana:2013sxa}.  However, we find it surprising that finite $w$ effects
merely alter the normalization and rapidity width of the proper energy, as opposed to changing its functional form altogether, 
and don't affect the fluid velocity.
This is especially noteworthy given that the $w$-dependence of the rapidity width $\xi_{\rm FWHM}$ in (\ref{eq:quadfits}) is not weak.
 
Additionally, we construct a simple framework to extend the universal initial hydrodynamic data for planar shocks to initial hydrodynamic data for
shocks with slowly varying transverse profiles.  Using this framework, we employ our planar shocks to 
study axisymmetric collisions.  
We  evolve our initial axisymmetric data forward in time
using viscous hydrodynamics and then compute the rapidity spectrum of produced hadrons using a Cooper-Frye freeze out prescription \cite{Cooper:1974mv}.  
We find that the rapidity distribution of massless particles is well approximated 
by a Gaussian with variance $1.9$ and $2.1$ for $\sqrt{s_{NN}}$ energies of 200 GeV and 2.76 TeV respectively.

An outline of our paper is as follows.
In Sec.~\ref{sec:setup} we develop the gravitational setup for our planar shock collisions. 
In Sec.~\ref{sec:resultsforplanar} we present our results for planar collisions, including early time non-universal transient effects,
and universal late time hydrodynamic evolution.
In Sec.~\ref{sec:transdyn} we develop the framework for including slowly varying transverse dynamics and present hydrodynamic simulations for axisymmetric collisions.
In Sec.~\ref{sec:spectrum} we present results for the spectrum of produced hadrons.  We conclude in Sec.~\ref{sec:conclusions}.

\section{Setup}
\label{sec:setup}

We construct initial data for Einstein's equations by superimposing the metric
of gravitational shock waves moving in the $\pm z$ directions.
In Fefferman-Graham coordinates,
the metric of a single shock moving in the $\pm z$ direction is
\begin{align}
    ds^2 &=  r^2 \big[
	{-} dt^2 + d\bm x_\perp^2 + dz^2 + {\textstyle \frac{dr^2}{r^4}}
	+ h_\pm(\bm x_\perp, z_\mp,r) \, dz_\mp^2
    \big] \,,
\label{eq:FG}
\end{align}
where $z_\mp \equiv z \mp t$, $\bm x_\perp$ are the two coordinates transverse to the collision axis, $r$ is the AdS radial coordinate, and
\begin{align}
    h_\pm(\bm x_\perp,z_\mp,r) &\equiv
     \int \frac{d^2 k}{(2\pi)^2} \>
    e^{i {\bm k} \cdot \bm x_\perp} \,
    \widetilde H_\pm({\bm k},z_\mp) \,
    \frac{8I_2(k/r)}{k^2 r^2} \,.
\label{eq:h}
\end{align}
The AdS curvature scale has been set to unity.
The boundary of the asymptotically AdS spacetime lies at radial coordinate $r = \infty$.
The single-shock metric (\ref{eq:FG}) is an exact solution to Einstein's 
equations for any choice of $\widetilde H_\pm$ \cite{Gubser:2008pc}.
This geometry represents a state in the dual SYM theory with stress tensor,
\begin{equation}
\label{eq:hatvsnohat}
T^{\mu \nu} = \frac{N_{\rm c}^2}{2 \pi^2} \widehat T^{\mu \nu},
\end{equation}
with non-zero components 
\begin{equation}
\label{eq:singleshock}
    \widehat T^{00} =  \widehat  T^{zz} =  \pm \widehat  T^{0z} = H_\pm(\bm x_\perp,z_\mp),
\end{equation}
where $H_\pm$ is the 2D transverse Fourier transform of $\widetilde H_\pm$.
Note that here and in what follows we used hats to denote quantities normalized by $\frac{N_{\rm c}^2}{2 \pi^2}$.
That is, for any function $F$ we define $\widehat F$ by $F = \frac{N_{\rm c}^2}{2 \pi^2} \widehat F$.

\begin{figure}[h]
\vskip +0.15in
\begin{center}
\includegraphics[scale = 0.45]{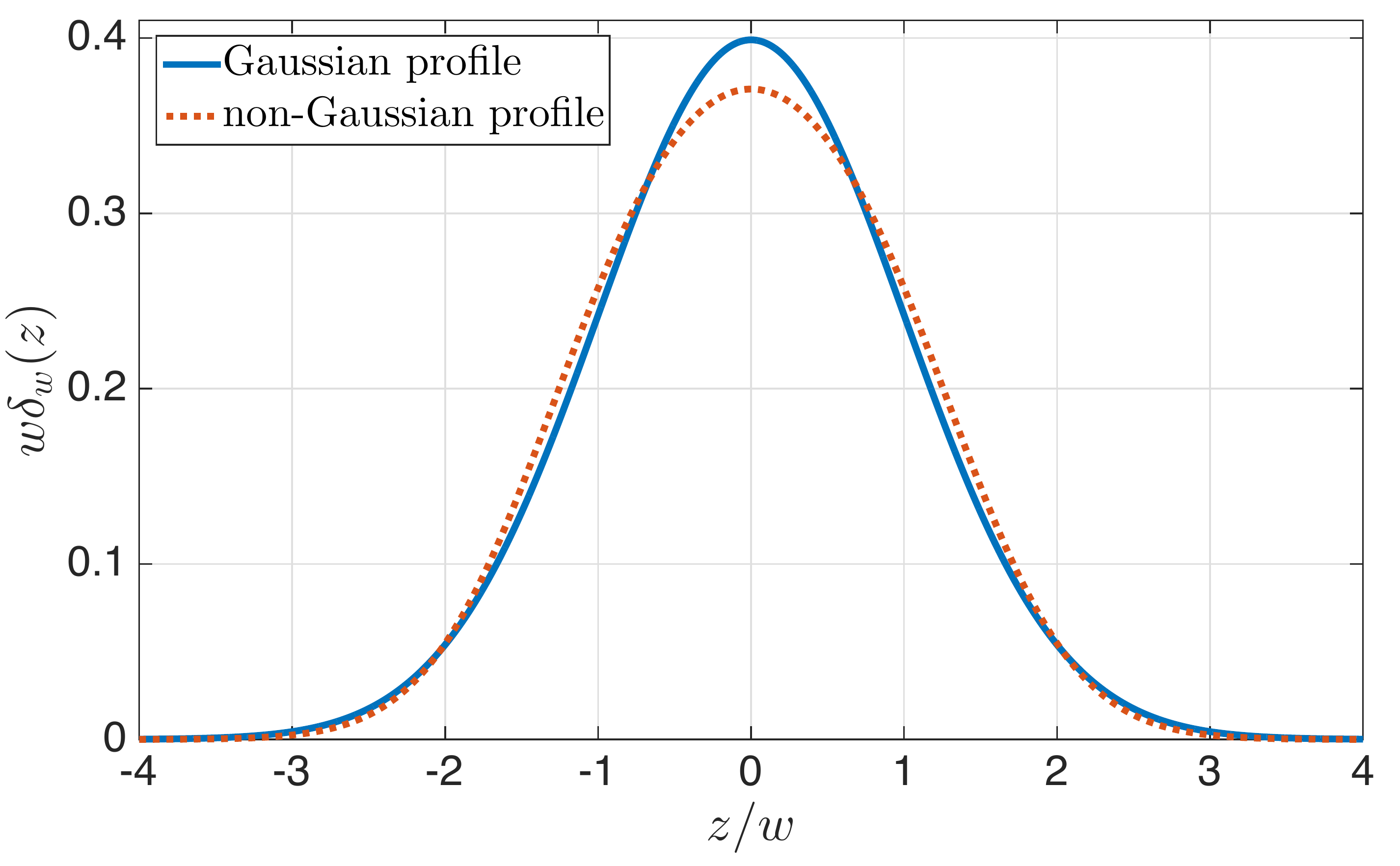}
\end{center}
\vskip -0.1in
\caption{%
The shock profiles given in Eq.~(\ref{eq:smearing}).  
\label{fig:smearingfunctions}
} 
\end{figure}

We choose 
\begin{equation}
\label{eq:initialH}
    H_\pm(\bm x_\perp,z_\mp)
    = \mu^3 \delta_w(z_\mp),
\end{equation}
where $\mu$ is an energy scale and $\delta_w(z_\mp)$ is a smeared $\delta$-function which satisfies
the normalization and variance conditions in 
Eq.~(\ref{eq:normandvar}).  We employ two different shock profiles
\begin{equation}
\label{eq:smearing}
\delta_w(z) =  \frac{1}{\sqrt{2 \pi w^2}} e^{-\frac 12 z^2/w^2} \ \ \ \ {\rm and} \ \ \ \ \delta_w(z) =  \frac{M}{e^{\frac 12 z^2/W^2} + 1},
\end{equation}
with
\begin{align}
M &= \textstyle 
\frac{1}{w} \sqrt{\textstyle-\frac{(4 + 3 \sqrt{2}) \zeta(\frac 32)}{ 4 \pi \zeta(\frac 12)^3} } \approx \frac{0.74}{w}, &
W &=  \textstyle 
w \sqrt{-\frac{\sqrt{2} \zeta(\frac 12)}{\zeta( \frac 32)}} \approx 0.89 w.
\end{align}
We refer to the first shock profile in (\ref{eq:smearing}) as the Gaussian profile
and the second as the non-Gaussian profile.  These shock profiles
are plotted in Fig.~\ref{fig:smearingfunctions}.

At early times, $t \ll -w$, the functions 
$h_\pm$ have negligible overlap and
the pre-collision geometry can be constructed from (\ref{eq:FG})
by replacing the last term with the sum of corresponding terms from
left and right moving shocks.
The resulting metric satisfies Einstein's equations, at early times,
up to exponentially small errors.

To evolve the pre-collision geometry forward in time through the collision we 
use the characteristic formulation of gravitational dynamics in
asymptotically AdS spacetimes discussed in detail in \cite{Chesler:2013lia}.
Our metric ansatz reads
\begin{equation}
    ds^2 = r^2 \, g_{\mu \nu}(x,r) \, dx^\mu dx^\nu + 2 \, dr \, dt \,,
\label{eq:ansatz}
\end{equation}
with Greek indices denoting spacetime boundary coordinates, $x^\mu = (t,\bm x_\perp,z)$.
Near the boundary,
$g_{\mu \nu} = \eta_{\mu \nu}  + g_{\mu \nu}^{(4)}/r^4 + O(1/r^5)$.
The sub-leading coefficients $g^{(4)}_{\mu\nu}$ determine 
the boundary stress tensor,
\begin{equation}
\label{eq:holostress}
\widehat T^{\mu \nu} = g_{\mu \nu}^{(4)}  + \tfrac{1}{4} \, \eta_{\mu \nu}\, g_{00}^{(4)}.
\end{equation}

To generate initial data for our characteristic evolution,
we numerically transform the pre-collision metric in Fefferman-Graham 
coordinates to the metric ansatz (\ref{eq:ansatz});
this requires computing a congruence of infalling radial null geodesics
and is outlined in \cite{vanderSchee:2014qwa}.
We then numerically solve Einstein's equations using the methods outlined in \cite{Chesler:2013lia}.
We measure  dimensional quantities in units of $\mu$.
For the Gaussian profile we study collisions with shock widths $w = n w_o$, $n = 1,2,\dots, 7$
where%
\footnote
  {
  For comparison,  Ref~\cite{Chesler:2010bi} used $w = 10 w_o$ and Ref~\cite{Casalderrey-Solana:2013aba} used $w$ in the range $0.66 w_o$ till $25 w_o$.
  }
\begin{equation}
w_o \equiv \frac{0.075}{\mu}.
\end{equation}
For the non-Gaussian profile we study collisions with $w = w_o$ and $w = 5 w_o$.
We time evolve all collisions $\mu t = 14$ units after the collision event.  After numerically solving Einstein's equations
we extract the boundary stress tensor via (\ref{eq:holostress}) and (\ref{eq:hatvsnohat}).

\section{Results for planar shocks}
\label{sec:resultsforplanar}

\subsection{Early time dynamics and non-universal transient effects}
\label{sec:earlytime}

\begin{figure}[ht]
\vskip -0.10in
\begin{center}
\includegraphics[scale = 0.5]{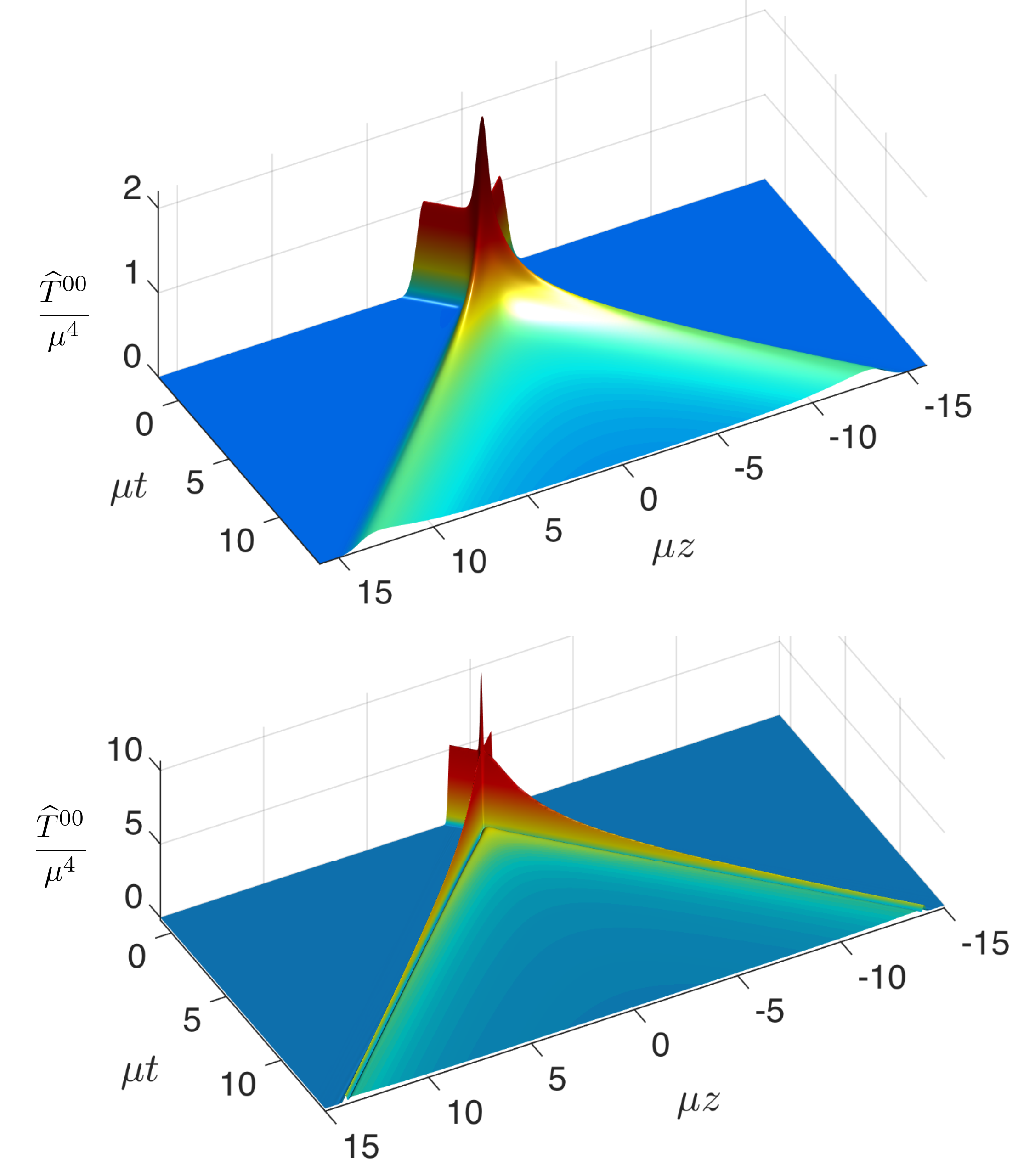}
\end{center}
\vskip -0.1in
\caption{%
Rescaled energy density $\widehat T^{00}$
as a function of time $t$ and longitudinal position $z$ for Gaussian shock profiles.
Top figure: wide shocks with width $w= 5 w_o$.
Bottom figure: narrow shocks with width $w = w_o$.
In both plots, the shocks approach each other along the $z$ axis
and collide at $z = 0$ at time $t = 0$.  The collisions
produce debris that fills the forward light cone.
Nevertheless, there are clear qualitative differences in the energy density near the forward light cone.
For $w = w_o$ there are clear post-collision remnants of the initial shocks propagating on the light cone.
These remnants decay with time like $t^{-p}$ with $p \approx 0.9$.%
\label{fig:shocksenergy}
} 
\end{figure}

\begin{figure}[h]
\vskip +0.15in
\begin{center}
\includegraphics[scale = 0.45]{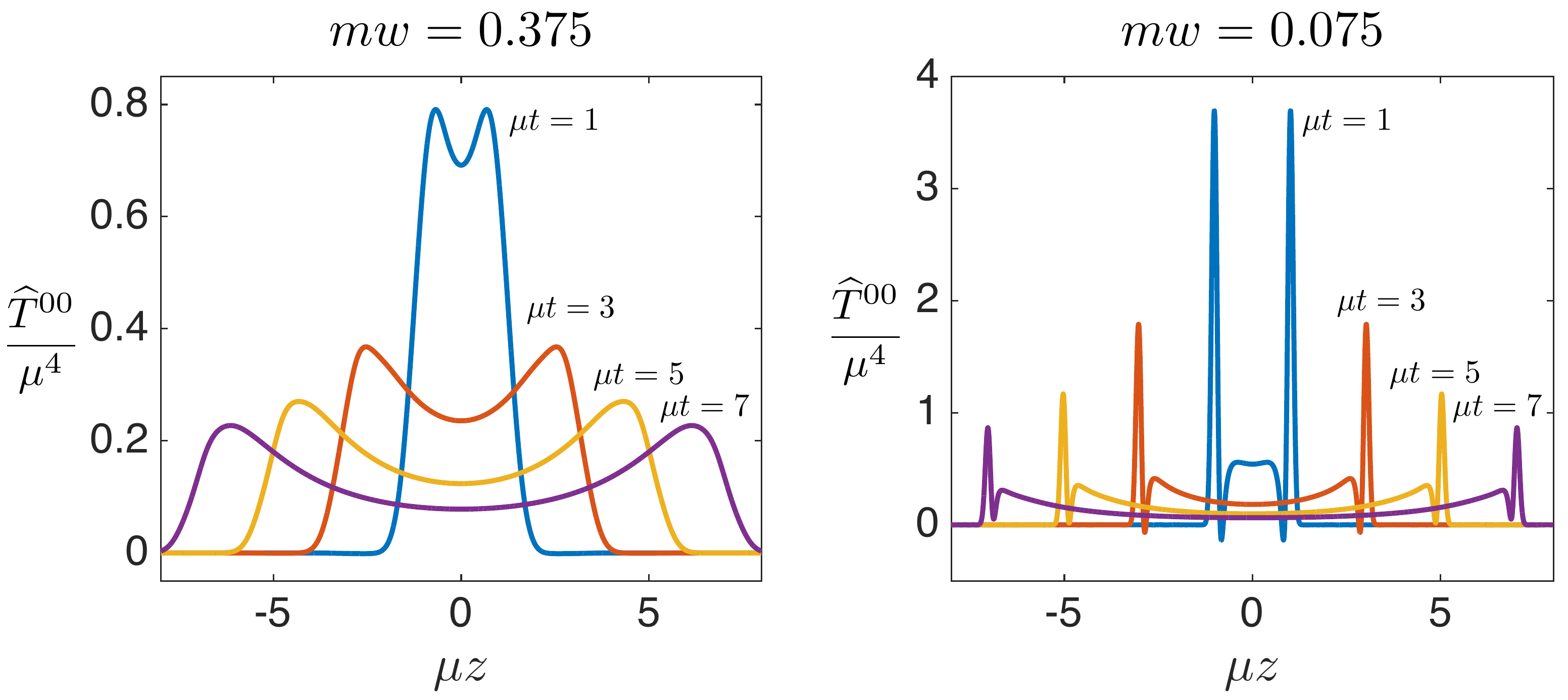}
\end{center}
\vskip -0.1in
\caption{%
Energy density $\widehat T^{00}/\mu^4$ at times $\mu t = 1,3,5,7$ for $w = 5 w_o$ (left)
and $w = w_o$ (right) for Gaussian shock profiles.  For $w = 5 w_o$ the post-collision energy density is smooth
with no distinct remnant of the shocks remaining on
the forward light cone.  In contrast, for $w = w_o$ 
there are clear remnants of the initial shocks propagating outward on the forward light cone.  
These remnants decay with time like $t^{-0.9}$.  Note the brief presence of negative energy behind the remnants 
in the $w = w_o$ collision.
\label{fig:energyatfixedtimes}
} 
\end{figure}

Let us begin by focusing on the energy density  produced by Gaussian shock collisions.
In Fig.~\ref{fig:shocksenergy} we plot the rescaled energy density $\widehat T^{00}$, 
for Gaussian shock profiles with widths $w = 5 w_o$ (top) and $w = w_o$ (bottom). 
The shocks approach each other at
the speed of light in the $\pm z$ direction and collide at $z = 0$ at time $t = 0$. For both cases,
the debris leaving the collision event appears dramatically different than the initial incoming
shocks.  Nevertheless, comparing the two plots in Fig.~\ref{fig:shocksenergy}, it is clear that 
there are qualitative differences in the energy density near the light cone.
To highlight these differences, in Fig.~\ref{fig:energyatfixedtimes} we plot 
$\widehat T^{00}$ at times $\mu t = 1,3,5,7$, again for Gaussian shock profiles with widths $w = 5 w_o$ and $w = w_o$.
As can be seen in both Figs.~\ref{fig:shocksenergy} and \ref{fig:energyatfixedtimes}, at width $w = w_o$ 
there are clear post-collision remnants of the shocks propagating outward on the light cone.  
In contrast, at width $w = 5w_o$ there are no signs of any distinct remnant of the shocks on
the forward light cone.  Instead, the post-collision energy density is smoothly distributed
in the interior of the forward light cone.
Moreover, for $w = 5 w_o$ the energy density is everywhere positive.  In contrast,
for $w = w_o$ one sees from 
Fig.~\ref{fig:energyatfixedtimes} the appearance of small regions of negative energy 
behind the receding maxima.  Clearly the behavior of the stress near the light cone
is sensitive to the shock width.

\begin{figure}[h]
\vskip +0.15in
\begin{center}
\includegraphics[scale = 0.45]{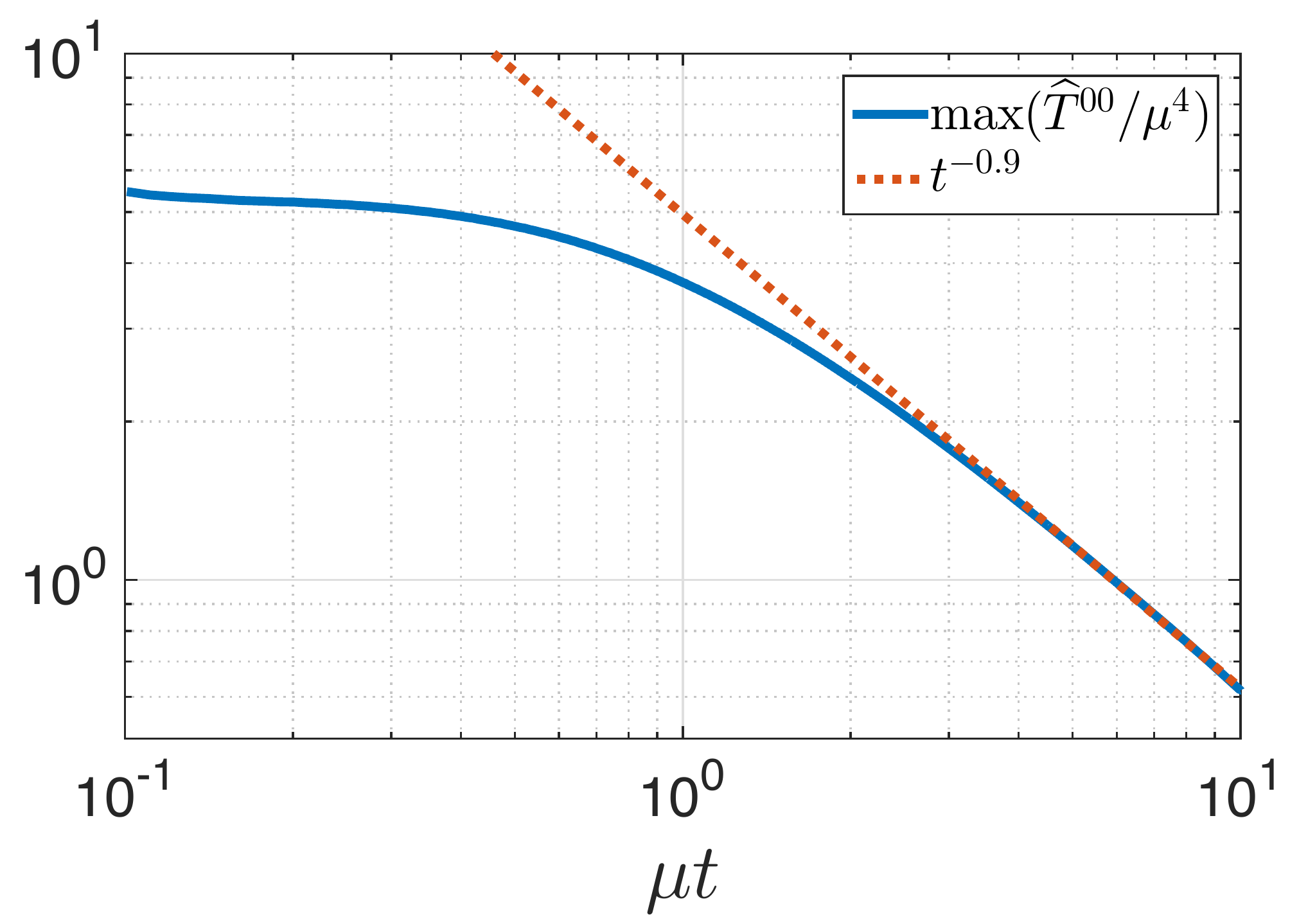}
\end{center}
\vskip -0.1in
\caption{%
The amplitude of the receding shock remnants as a function of time for Gaussian shock profile with width $w = w_o$.  
Our numerics are consistent with a power law decay $t^{-p}$ with $p \approx 0.9$
\label{fig:decayofmaxima}
} 
\end{figure}

However, both the presence of negative energy and shock remnants 
on the forward light cone in the $w=w_o$ collision are transient effects.  Indeed, from 
Fig.~\ref{fig:energyatfixedtimes} we see that already by time $\mu t = 5$ the regions of negative energy have disappeared.
In Fig.~\ref{fig:decayofmaxima}
we plot the amplitude of the outgoing energy maxima for width $w = w_o$ as a function of time.
Our numerics are consistent with a $t^{-0.9}$ power law decay, which was also seen in \cite{Chesler:2013lia}.
Note that a power law decay is natural in a conformal theory, where there is no intrinsic scale.
By time $\mu t = 14$ more than 90\% of the initial shock energy lies in the forward light cone.
Hence, our results suggest that irrespective of the shock width, the collision results in the complete
annihilation of the shocks with essentially
all energy lying well inside the forward light cone at late times.

\begin{figure}[h]
\vskip +0.20in
\begin{center}
\includegraphics[scale = 0.45]{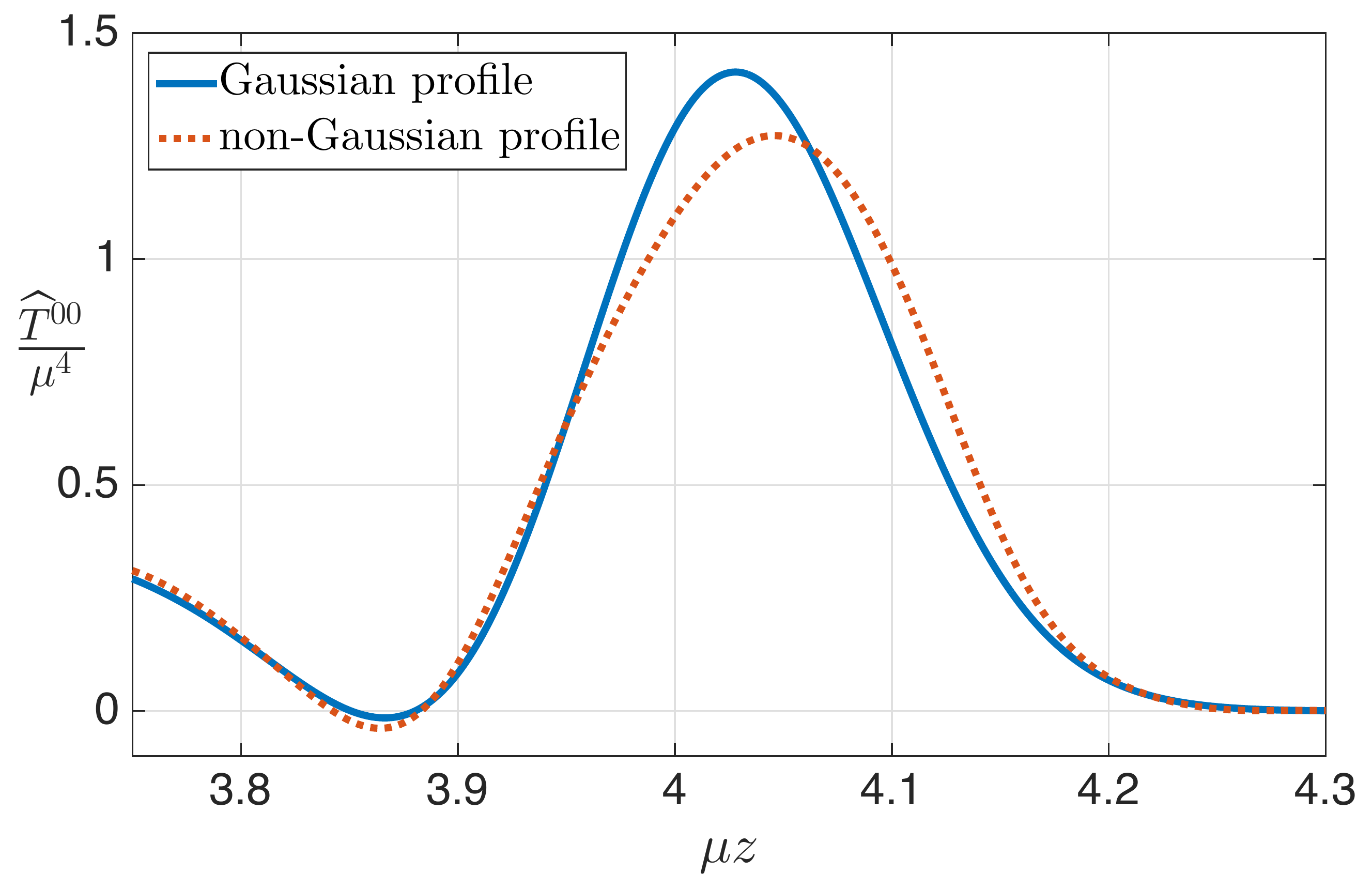}
\end{center}
\vskip -0.1in
\caption{%
The energy density at $\mu t = 4$ with $w = w_o$ for both Gaussian and non-Gaussian shock profiles.
The size of the difference in the energy densities should be compared to that of the different pre-collision shock profiles
shown in Fig.~\ref{fig:smearingfunctions}.  Clearly, near the light cone the stress is sensitive to the structure of the shock profiles.
\label{fig:nonuniversalnearlightcone}
} 
\end{figure}

How sensitive is the stress near the forward light cone to perturbations in the functional form of the shock profile?  
In Fig.~\ref{fig:nonuniversalnearlightcone} we plot the energy density at time $\mu t = 4$ 
with width $w = w_o$ for both Gaussian and non-Gaussian shock profiles.
Clearly, near the light cone the stress is different for the different shock profiles.
The relative magnitude of the change should be compared to that of the different 
profiles shown in Fig.~\ref{fig:smearingfunctions}.  Evidently, in addition to being sensitive to the width of the shocks, 
the stress tensor near the future
light cone is also sensitive to the functional form of the shock profile.
We note, however,  that the amplitude of the 
decaying shock remnants also scales like $t^{-0.9}$ for the non-Gaussian profile.

\subsection{Universal initial data for hydrodynamics}
According to fluid/gravity duality \cite{Bhattacharyya:2008jc,Baier:2007ix},
at sufficiently late times
the evolution of the stress tensor should be governed by hydrodynamics.
How do the decaying shock remnants and negative energy density near the light cone
imprint themselves on the hydrodynamic evolution?  Are there qualitative differences between the 
hydrodynamic evolution for thick and thin shocks? To address these questions we first identify the 
domain $\mathcal R$ in spacetime in which hydrodynamics is a good description of the evolution of the stress and then study
the $w$ and $\delta_w$ dependence of the hydrodynamic variables on a fixed Cauchy surface in $\mathcal R$.

In relativistic neutral fluid hydrodynamics the hydrodynamic variables are typically taken to be the proper energy density
$\epsilon$ and the fluid velocity $u^\mu$.   The fluid velocity is defined 
to be the normalized time-like $(u_\mu u^\mu = -1)$ future directed $(u^0 > 0)$
eigenvector of the stress tensor,
\begin{equation}
T^{\mu}_{\ \nu} \, u^\nu = -\epsilon \, u^\mu \,,
\label{eq:veldef}
\end{equation}
with $\epsilon$ the associated eigenvalue.   
In terms of $\epsilon$ and $u^\mu$
the constitutive relations of fluid/gravity read 
\begin{equation}
\label{eq:hydroconstituative}
T^{\mu \nu}_{\rm hydro} = p g^{\mu \nu} + (\epsilon + p) u^\mu u^\nu + \Pi^{\mu \nu},
\end{equation}
where $p = \frac{\epsilon}{3}$ is the pressure and $\Pi^{\mu \nu}$ is the 
viscous stress.  The viscous stress satisfies $u_\mu \Pi^{\mu \nu} = 0$
and $g_{\mu \nu} \Pi^{\mu \nu} = 0$ and at first order in gradients is 
given by $\Pi^{\mu \nu} = - \eta \sigma^{\mu \nu}$ with
$\eta$ the shear viscosity and
\begin{equation}
\sigma_{\mu \nu} = \partial_{(\mu} u_{\nu )} + u_{(\mu} u^\rho \partial_\rho u_{\nu)} - {\textstyle \frac{1}{3}} \partial_\alpha u^\alpha \left [ \eta_{\mu \nu} + u_\mu u_\nu \right],
\end{equation}
the shear tensor.   The shear viscosity may be expressed in terms of the proper energy via \cite{Kovtun:2004de,Policastro:2001yc}
\begin{equation}
\label{eq:viscosity}
\eta = \frac{1}{3 \pi T} \epsilon,
\end{equation}
with the temperature $T$ given by 
\begin{equation}
\label{eq:Tdef}
T = \left (\frac{8 \epsilon}{ 3 \pi^2 N_{\rm c}^2} \right )^{1/4}.
\end{equation}
The hydrodynamic equations of motion are given by the energy-momentum 
conservation equation $\partial_\mu T^{\mu \nu}_{\rm hydro} = 0$.
Note that the hydrodynamic stress tensor is completely determined by four functions.
In contrast, in general the exact (traceless) stress tensor contains nine independent function.

Instead of solving the hydrodynamic equations of motion for the evolution of $\epsilon$
and $u^\mu$, a simple way to compare the gravitational evolution to hydrodynamics is to
extract the exact $\epsilon$ and $u^\mu$ from the eigenvalue equation (\ref{eq:veldef}) with the exact stress tensor.
In the domain where hydrodynamics is a good description this should yield the same time evolution for
proper energy density and fluid velocity as hydrodynamics.
With the exact proper energy and fluid velocity known, we can then construct $T^{\mu \nu}_{\rm hydro}$
from Eqs.~(\ref{eq:hydroconstituative})--(\ref{eq:Tdef}) and compare $T^{\mu \nu}$ and $T^{\mu \nu}_{\rm hydro}$.
To quantify the domain in which hydrodynamics is applicable we then
define the residual measure 
\begin{equation}
\label{eq:deltadef}
\Delta \equiv \frac{1}{ p}\sqrt{\Delta T_{\mu \nu}\Delta T^{\mu \nu}}, \ \ \  \ \ \ \Delta T^{\mu \nu} \equiv T^{\mu \nu} - T^{\mu \nu}_{\rm hydro}.
\end{equation}
The quantity $\Delta$,
evaluated in the local fluid rest frame,
measures the relative difference between the spatial stress in
$T^{\mu \nu}$ and $T^{\mu \nu}_{\rm hydro}$.
Regions of spacetime with $\Delta \ll 1$ are evolving hydrodynamically.

\begin{figure}[h]
\vskip 0.15in
\begin{center}
\includegraphics[scale = 0.4]{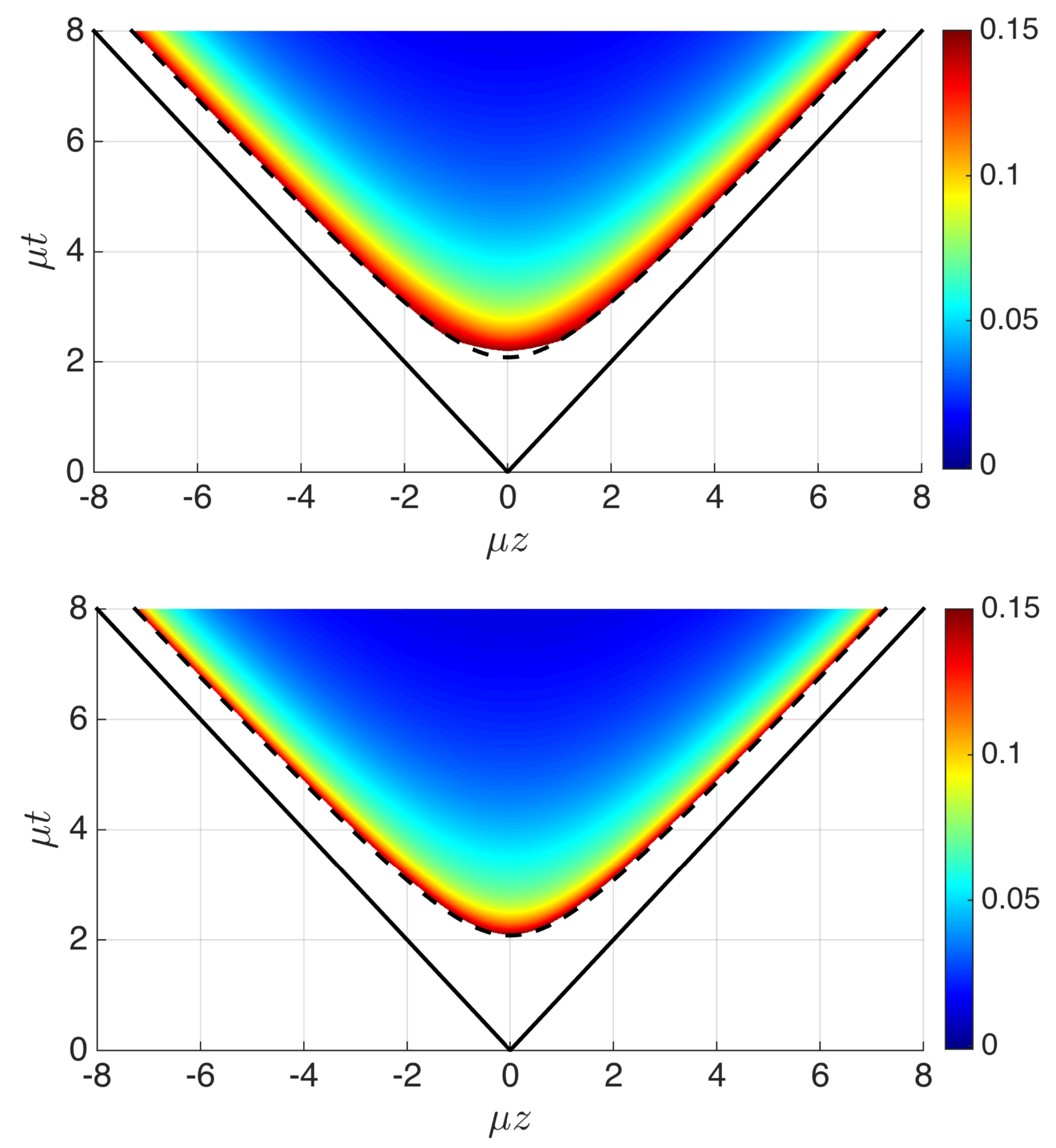}
\end{center}
\vskip -0.1in
\caption{%
The hydrodynamic residual $\Delta$ defined in Eq.~(\ref{eq:deltadef})
for Gaussian shock profiles with widths $w = 5 w_o$ (top) and $w = w_o$ (bottom).  Regions with 
$\Delta \ll 1$ are well described by viscous hydrodynamics.  
At time $\mu t = 8$ the minimum values of $\Delta$ are $0.015$ and $0.013$ for
$w = 5 w_o$ and $w = w_o$ respectively.  
Note that we have plotted $\Delta$ only in the region $\Delta \leq 0.15$.   
This region is bounded by the dashed curves (\ref{eq:hydrodomain}), which are the same for both collisions.
\label{fig:hydroresidual}
} 
\end{figure}

Let us first focus on the hydrodynamic evolution produced by Gaussian shock collisions.
In Fig.~\ref{fig:hydroresidual} we plot the hydrodynamic residual $\Delta$ for Gaussian shock profiles
with widths $w = 5 w_o$ (top) and $w = w_o$ (bottom).
Note that we only plot $\Delta$ in the region $\mathcal R$ defined to be the largest connected region in spacetime where $\Delta \leq 0.15$.
Outside of $\mathcal R$ hydrodynamics is not a good description and the fluid velocity need not even be well defined 
\cite{Arnold:2014jva} (\textit{i.e.} the stress need not have a time-like eigenvector).
The dashed line in the figure, which bounds the region $\mathcal R$, is given by
\begin{equation}
\label{eq:hydrodomain}
\tau_* = \sqrt{(t - \Delta t)^2 -z^2},
\end{equation}
with $\mu \tau_* = 1.5$ and $\mu \Delta t = 0.58$.
We therefore conclude that the domain of applicability of hydrodynamics is approximately the same for both shock thicknesses.
Fig.~\ref{fig:hydroresidual} 
clearly shows that our planar shock collisions result in the formation of an expanding volume of fluid
which is well described by hydrodynamics everywhere except near the light cone, where non-hydrodynamic 
effects become important.  At mid-rapidity viscous hydrodynamics
becomes a good approximation at time 
\begin{equation}
\mu t_{\rm hydro} \approx 2.
\end{equation}

With the applicability of hydrodynamics established, we now turn to the $w$ dependence of the initial 
hydrodynamic data.  We introduce proper time $\tau$ and rapidity $\xi$ coordinates via
\begin{align}
t &= \tau \cosh \xi,
&z & = \tau \sinh \xi,
\end{align}
and study the hydrodynamic variables $u^\mu$ and $\epsilon$ on the 
$\tau = \tau_{\rm init}$ Cauchy surface with 
\begin{equation}
\mu \tau_{\rm init} = 3.5.
\end{equation}
Note that in what follows
we restrict the rapidity range to that bounded by Eq.~(\ref{eq:hydrodomain}), where 
$\Delta \leq 0.15$ and hydrodynamics is a good description.

\begin{figure}[h]
\vskip 0.15in
\begin{center}
\includegraphics[scale = 0.36]{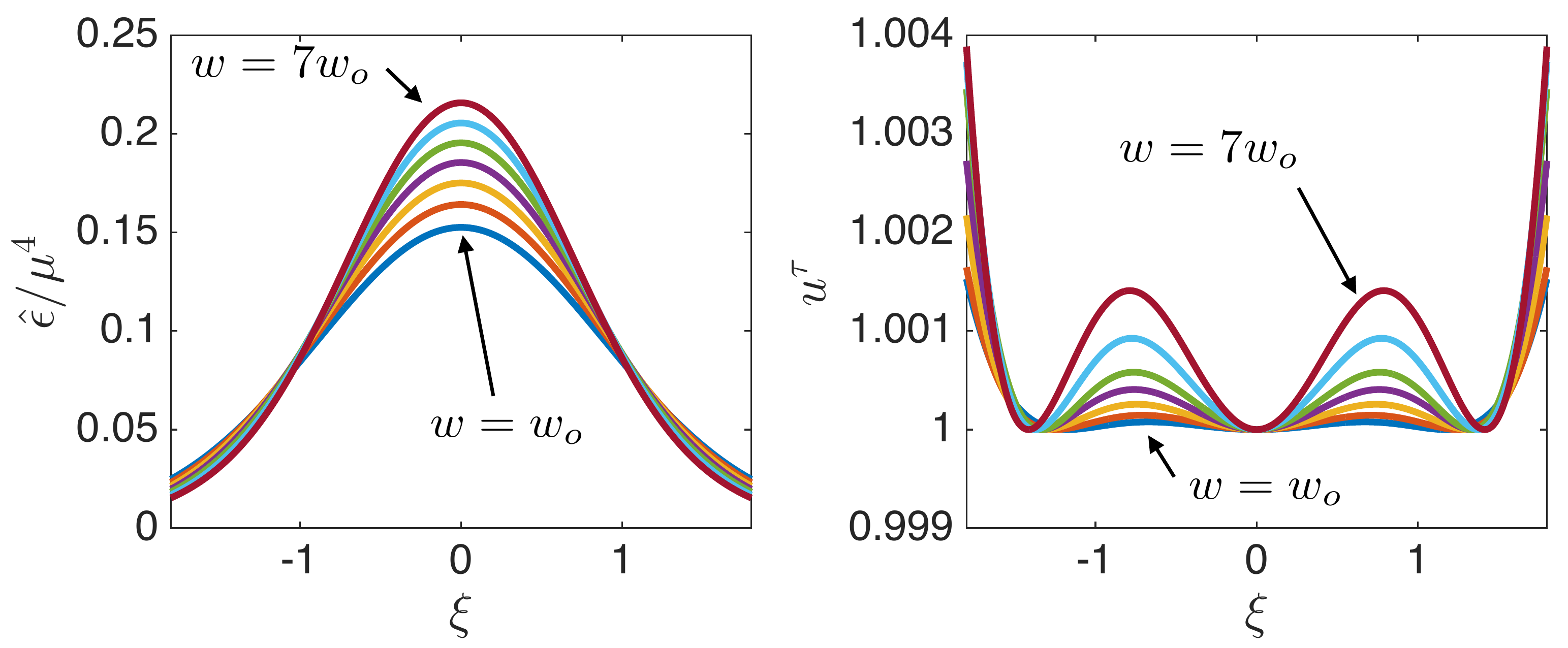}
\end{center}
\vskip -0.1in
\caption{%
The proper energy (left) and the proper time component of the 
fluid velocity (right) as a function of rapidity $\xi$ at proper time $\mu \tau_{\rm init} = 3.5$
for Gaussian shock profiles with widths $w = n w_o$, $n = 1,2,\dots,7$.   
Note that in units of $\mu$ the range of $w$ shown is $\mu$ is $\mu w = 0.075$ to $\mu w = 0.525$.  
The fluid velocity is well described by boost invariant flow with $u^\tau \approx 1$.
Note that 
we restrict the rapidity range to that bounded by Eq.~(\ref{eq:hydrodomain}), where 
$\Delta \leq 0.15$ and hydrodynamics is a good description.
\label{fig:initialhydrodata}
} 
\end{figure}

\begin{figure}[h]
\vskip 0.15in
\begin{center}
\includegraphics[scale = 0.355]{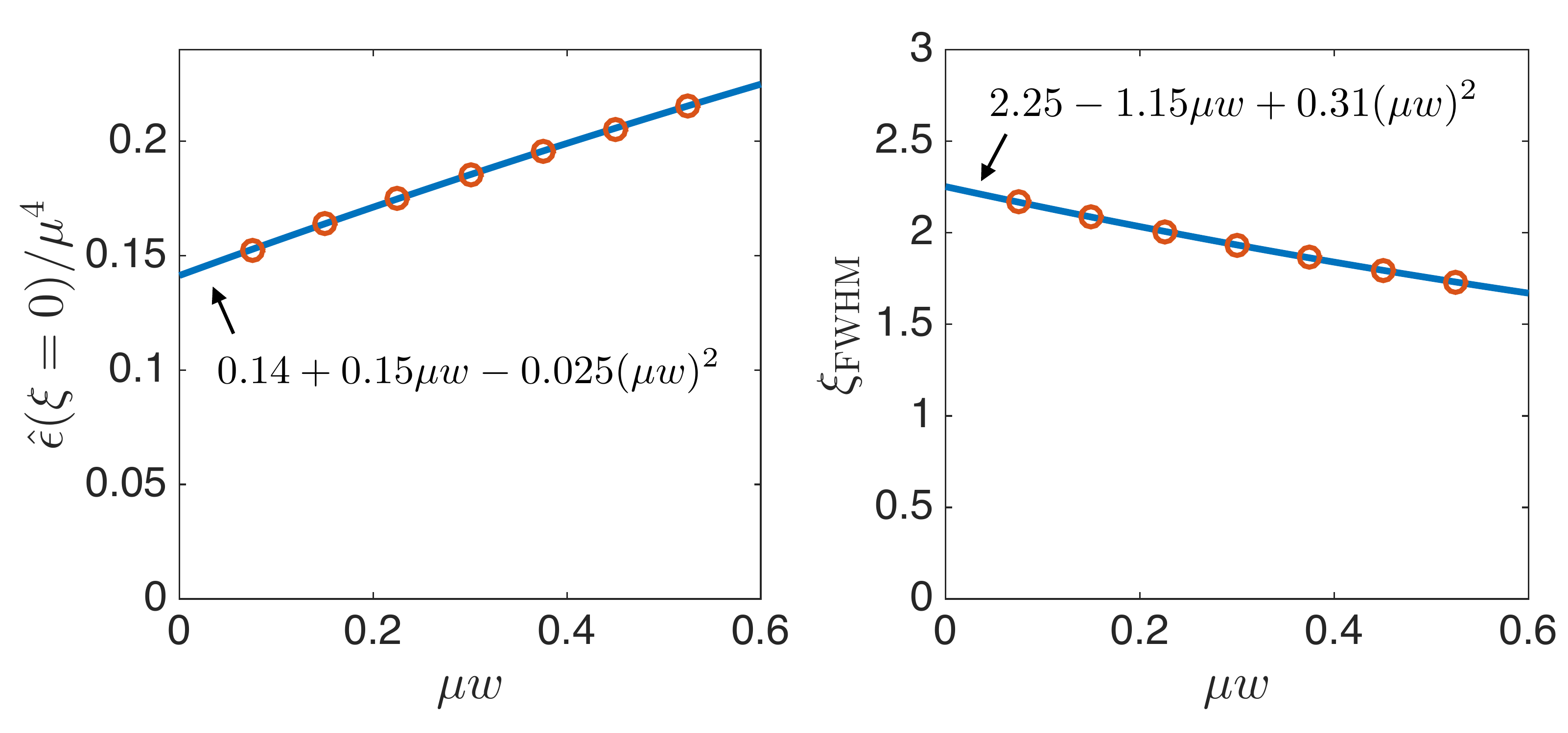}
\end{center}
\vskip -0.1in
\caption{%
Left: the normalized proper energy at rapidity $\xi = 0$ as a function of shock width.  
Right: the full width half max rapidity of the proper energy density as a function of shock width. 
Both plots are at proper time  $\tau = \tau_{\rm init}$ and are for Gaussian shock profiles.
Also included in both plots are the quadratic fits (\ref{eq:quadfits}).
\label{fig:fwhm}
} 
\end{figure}

In Fig.~\ref{fig:initialhydrodata} we plot $\epsilon$ and the $\tau$-component of the fluid velocity, $u^\tau$,
as a function of rapidity at $\tau = \tau_{\rm init}$ for Gaussian shock profiles.  Included in the figure
are shock widths $w = n w_o$, $n = 1,2,\dots,7$.  
Recall that for boost invariant flow $u^\tau = 1$.
We see from the figure that $u^\tau \approx 1$ with narrower shocks 
having $u^\tau$ closer to $1$ than wider shocks.  
We therefore conclude that for all shock widths shown the initial fluid velocity is very well
described by boost invariant flow.  Turning to the proper energy density, we see that thinner shocks lead to 
a broader rapidity profile with smaller amplitude than wider shocks.  
In Fig.~\ref{fig:fwhm} we plot the amplitude $\epsilon(\xi = 0)$ and the full width at half maximum rapidity $\xi_{\rm FWHM}$
of $\epsilon$ as a function of shock thickness $w$, again for Gaussian shock profiles.  
Also included in the plots are the quadratic fits in Eqs.~(\ref{eq:quadfits}),
which clearly well-describe the numerical data.
Note  $\frac{d\xi_{\rm FWHM}}{d (\mu w)} \approx -1$, indicating appreciable $w$ dependence in the 
rapidity width.

\begin{figure}[h]
\vskip 0.15in
\begin{center}
\includegraphics[scale = 0.36]{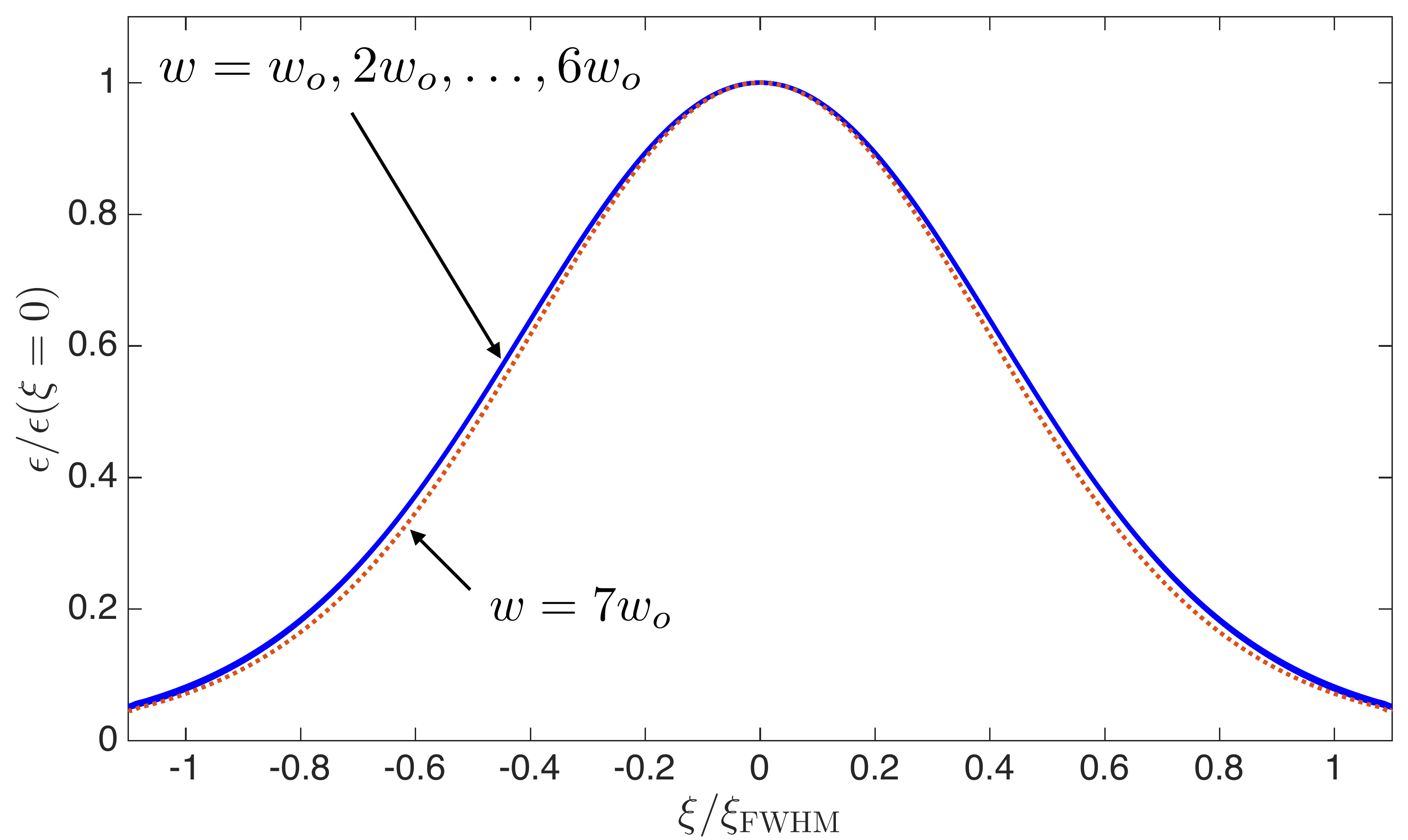}
\end{center}
\vskip -0.1in
\caption{%
The rescaled proper energy density $\epsilon/\epsilon(\xi = 0)$  for Gaussian shock profiles at fixed $\tau = \tau_{\rm init}$ 
as a function of the rescaled rapidity $\xi/\xi_{\rm FWHM}$.  Note that in units of $\mu$ the range of $w$ shown is $\mu$ is $\mu w = 0.075$ to $\mu w = 0.525$.  
When rescaled, all proper energy curves seen in Fig.~\ref{fig:initialhydrodata}
collapse onto each other.  
\label{fig:rescaledproperenergy}
} 
\end{figure}

A striking feature of the initial hydrodynamic data presented in Fig.~\ref{fig:initialhydrodata}
is the absence of any distinct qualitative change in either the proper energy or fluid velocity as 
the shock thickness is varied from $w = 7 w_o$ to $w = w_o$.  Indeed, there is very little quantitative change
in $u^\tau$ as the shock thickness is varied from $w = 7 w_o$ to $w = w_o$.  
This stands in stark contrast to the behavior of the stress near the forward light cone,
which changes qualitatively as the shock width is varied.  
In Fig.~\ref{fig:rescaledproperenergy} we plot the normalized proper energy $\epsilon/\epsilon(\xi = 0)$ 
at proper time $\tau_{\rm init}$
as a function of normalized rapidity $\xi/\xi_{\rm FWHM}$ for Gaussian shock profiles.  Included in the figure
are shock widths $w = n w_o,$ $n = 1,2,\dots,7.$
Remarkably, when rescaled all the proper energy curves in Fig.~\ref{fig:rescaledproperenergy}
collapse onto one single curve!   Only at $w = 7 w_o$ do we see a small discrepancy between the different curves. 
This observation implies that the initial hydrodynamic proper energy has the form
of Eq.~(\ref{eq:gaussianform0}),
with all $w$ dependence solely in the normalization and rapidity width
of the proper energy.

\begin{figure}[h]
\vskip 0.15in
\begin{center}
\includegraphics[scale = 0.45]{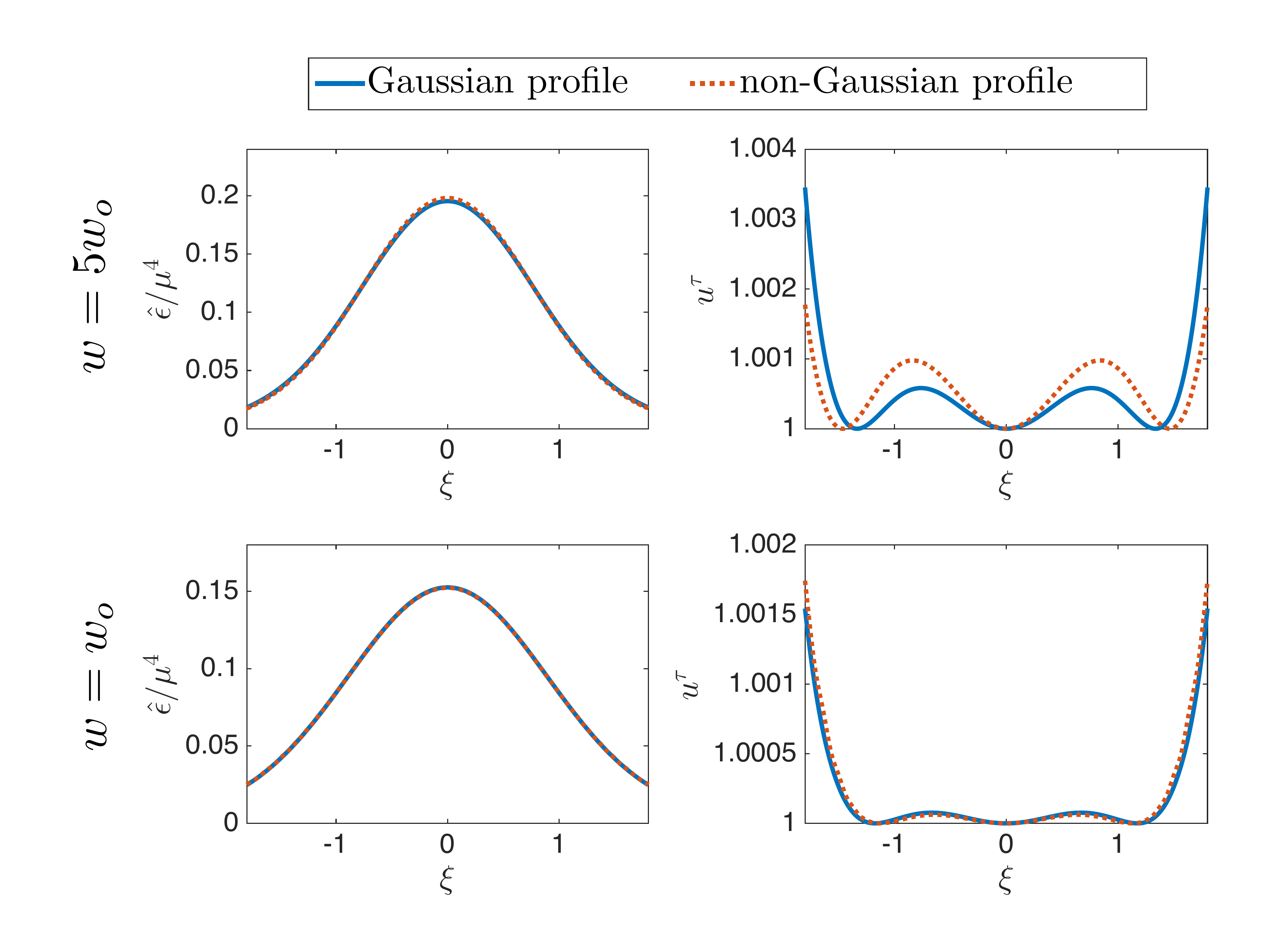}
\end{center}
\vskip -0.1in
\caption{%
The proper energy (left) and $\tau$-component of the fluid velocity, $u^\tau$, as a function of rapidity $\xi$ at proper time $\tau = \tau_{\rm init}$
for both Gaussian and non-Gaussian shock profiles with widths $w = 5 w_o$ (top) and $w = w_o$ (bottom).  At both shock thicknesses we see that 
both the proper energy and fluid velocity
are insensitive to the choice of shock profile.\label{fig:universalhydro}
} 
\end{figure}

Is the hydrodynamic evolution sensitive to perturbations in the shock profile?  
To answer this question, in Fig.~\ref{fig:universalhydro} we plot $\hat \epsilon$ and $u^\tau$
again at $\tau = \tau_{\rm init}$ with widths $w = 5w_o$ (top) and $w = w_o$ (bottom) for both Gaussian and 
non-Gaussian shock profiles.  As is evident from the figure, both $\epsilon$ and $u^\tau$ 
are nearly identical for both shock profiles.  This should be contrasted with Fig.~\ref{fig:nonuniversalnearlightcone},
where the energy density near the light cone was seen to be sensitive to the 
shock profile.  Evidently, the initial hydrodynamic data is insensitive to the precise 
functional form of the shock profile $\delta_w$.

\begin{figure}
\vskip 0.15in
\begin{center}
\includegraphics[scale = 0.4]{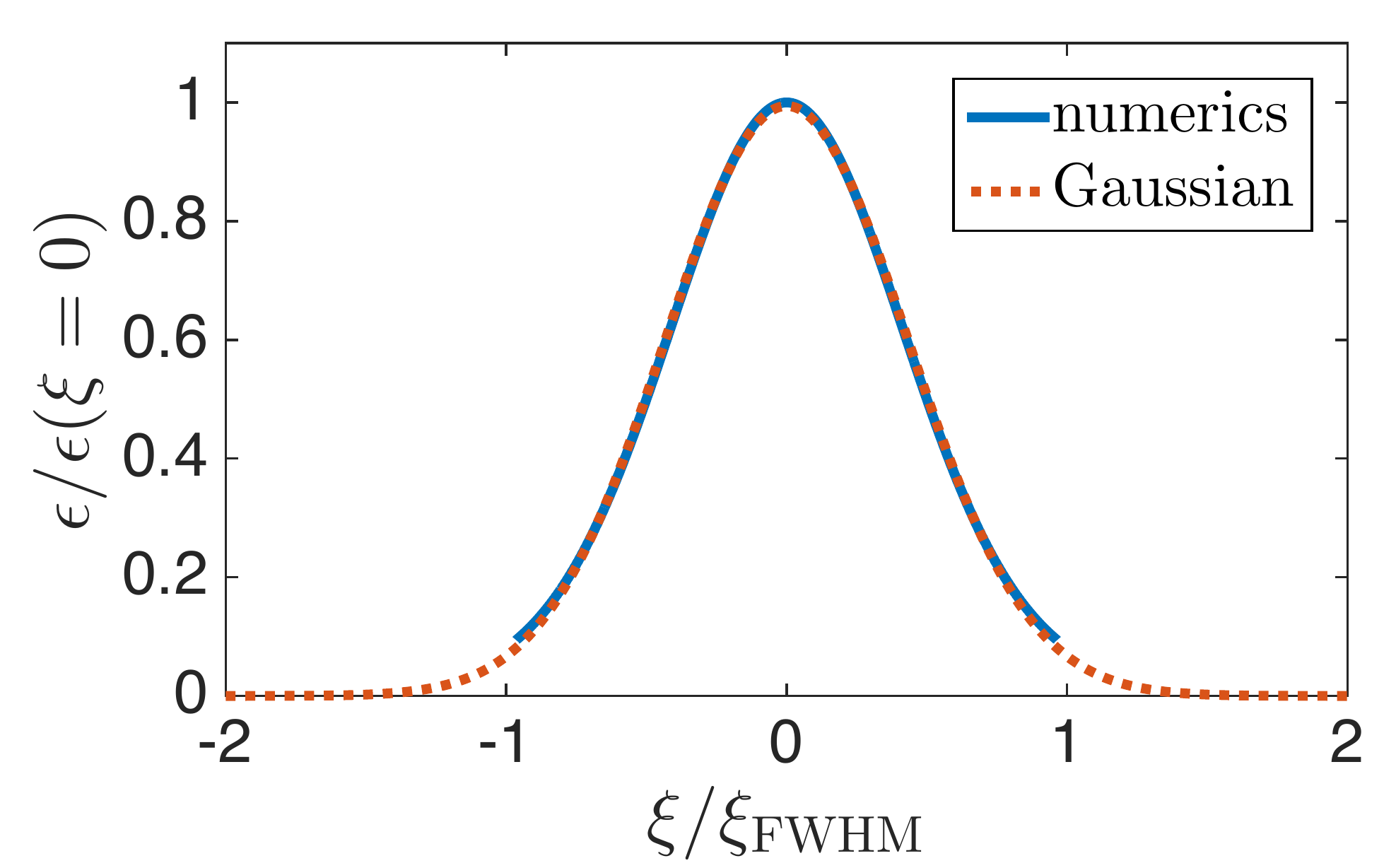}
\end{center}
\vskip -0.1in
\caption{%
The rescaled proper energy at $\tau = \tau_{\rm init}$ as a function of the rescaled rapidity $\xi/\xi_{\rm FWHM}$.  Also included is the Gaussian (\ref{eq:unitgaussian}),
which has unit full width at half maximum. 
\label{fig:properenergywithgaussian}
} 
\end{figure}

What is the function $f$ in Eq.~(\ref{eq:gaussianform0})?  
In Fig.~\ref{fig:properenergywithgaussian} we plot $\epsilon$ as a function of $\xi/\xi_{\rm FWHM}$ for $w = 5 w_o$ together with the Gaussian
(\ref{eq:unitgaussian}),
which has unit full width at half maximum.  Evidently, 
the initial hydrodynamic data is well described by a boost invariant fluid velocity and a Gaussian proper energy rapidity profile.

While we have restricted our numerical analysis to $w_o \leq w \leq 7 w_o$, we note that 
we see no evidence of the above universal hydrodynamic behavior disappearing as 
$w$ is further decreased.  Why?  First of all, in Fig.~\ref{fig:rescaledproperenergy} we see no sign that the functional form (\ref{eq:gaussianform0})
of the proper energy changes as $w$ is decreased.
Second, as shown in Fig.~\ref{fig:fwhm}, both the width and amplitude of the proper energy 
are well-described by the quadratic fits (\ref{eq:quadfits}) and show no signs of additional structure at small $w$.
Last, as shown in Fig.~\ref{fig:initialhydrodata}, it appears that boost invariant flow becomes a better and better approximation 
to the initial fluid velocity as $w$ decreases.   A natural interpretation of these observations is that the produced hydrodynamic flow has a smooth 
$w \to 0$ limit, in which the incoming shocks become $\delta$-functions, with the initial proper energy being a Gaussian in rapidity and the initial fluid velocity given by boost 
invariant flow.  Using the fits (\ref{eq:quadfits}), we extrapolate to $w = 0$ and estimate the initial width and amplitude of the proper energy for $\delta$-function collisions to be
$\xi_{\rm FWHM}|_{w = 0} \approx 2.25$ and $\hat \epsilon(\xi = 0)|_{w = 0}/\mu^4 \approx 0.14$.

\section{Including transverse dynamics during hydrodynamic evolution}
\label{sec:transdyn}

Heavy ion collisions are of course not translationally invariant as our planar shock collisions are.
Let us for simplicity focus on head-on collisions with zero impact parameter.  
A gravitational model of a heavy ion with non-trivial transverse profile is the shock metric (\ref{eq:FG})
with the function $H_{\pm}$ given by 
\begin{equation}
\label{eq:HFull}
H_\pm(\bm x_{\perp},z_{\mp}) =  \frac{\widehat E}{V} \left [ 1 + \exp \left ( \textstyle \frac{\sqrt{\bm x_{\perp}^2 + {\textstyle \frac{1}{\gamma^2}} z_\mp^2} - R}{a} \right) \right ]^{-1},
\end{equation}
with the constant $V$ fixed by the condition $\int dz d^2 x_\perp H_\pm = \widehat E$ (so $E = \frac{N_{\rm c}^2}{2 \pi^2} \widehat E$ is the total shock energy).  
The function (\ref{eq:HFull})
is simply a Woods-Saxon potential translating in the $\pm z$ direction at the speed of light.
The parameter $R$ is the nuclear radius and the parameter $a$
is the nuclear surface thickness.   The parameter $\gamma$ mimics the effects of Lorentz contraction in the $z$-direction.

In the limit where transverse gradients are small, at each $\bm x_\perp$ the stress tensor must be that of planar shock 
collisions.   This happens when the nuclear radius $R$ is much greater than the hydrodynamization time $t_{\rm hydro}$.
Therefore, when $R \gg t_{\rm hydro}$ we can construct initial hydrodynamic data at some early time merely using 
planar shock collisions.  The future evolution --- including transverse dynamics --- can then be studied using hydrodynamics.
To this end we define the $\bm x_\perp$-dependent 
energy scale $\mu(\bm x_\perp)$ and longitudinal width $w(\bm x_\perp)$ via
\begin{align}
\mu(x_\perp)^3 &\equiv \int dz \, H_\pm(\bm x_{\perp},z_{\mp}), &
w(\bm x_\perp)^2 &\equiv \frac{\int dz \, z^2 H_\pm(\bm x_{\perp},z_{\mp})}{\int dz \, H_\pm(\bm x_{\perp},z_{\mp})}.
\end{align}
As we shall see below, for energies at RHIC and the LHC the local width $\mu(\bm x_\perp) w(\bm x_\perp) \lesssim 1/2$ and the initial hydrodynamic data 
falls within the domain of universality seen above in Figs.~\ref{fig:initialhydrodata} and \ref{fig:rescaledproperenergy}.   In 
other words, the initial hydrodynamic data at some $\bm x_\perp$ only depends on the local energy scale $\mu(\bm x_\perp)$
and the local width $w(\bm x_\perp)$ and not on the precise longitudinal structure of the 
shock profile (\ref{eq:HFull}).

Let us henceforth denote the post-collision stress tensor for planar collisions
by $T^{\mu \nu}_{\rm planar}$. 
 $T^{\mu \nu}_{\rm planar}$ can be written
\begin{equation}
T^{\mu \nu}_{\rm planar}(\tau,\xi,w) = \mu^4 \mathcal T^{\mu \nu}_{\rm planar}(\mu \tau, \xi,\mu w),
\end{equation}
where $\mathcal T^{\mu \nu}_{\rm planar}(\cdot,\cdot,\cdot)$ is a dimensionless function
of three dimensionless arguments and is independent of the structure of the colliding shocks.
Therefore, in the limit where transverse gradients are small we can write the stress tensor 
as
\begin{equation}
\label{eq:hydroIC}
T^{\mu \nu}(\tau,\bm x_\perp,\xi) =  \mu(\bm x_\perp)^4 \mathcal T^{\mu \nu}_{\rm planar}( \mu(\bm x_\perp) \tau,\xi , \mu(\bm x_\perp) w(\bm x_\perp)).
\end{equation}
Eq.~(\ref{eq:hydroIC}) is valid for times $\tau \ll \ell$ with $\ell$ the typical length scale over which $ \mu(\bm x_\perp)$ varies.
Since the local hydrodynamization time is
of order $1/ \mu$, we may use (\ref{eq:hydroIC}) to construct initial data for hydrodynamics when 
\begin{equation}
\label{eq:sepofscales}
\ell \mu \gg 1.
\end{equation}

\begin{figure}
\vskip 0.15in
\begin{center}
\includegraphics[scale = 0.45]{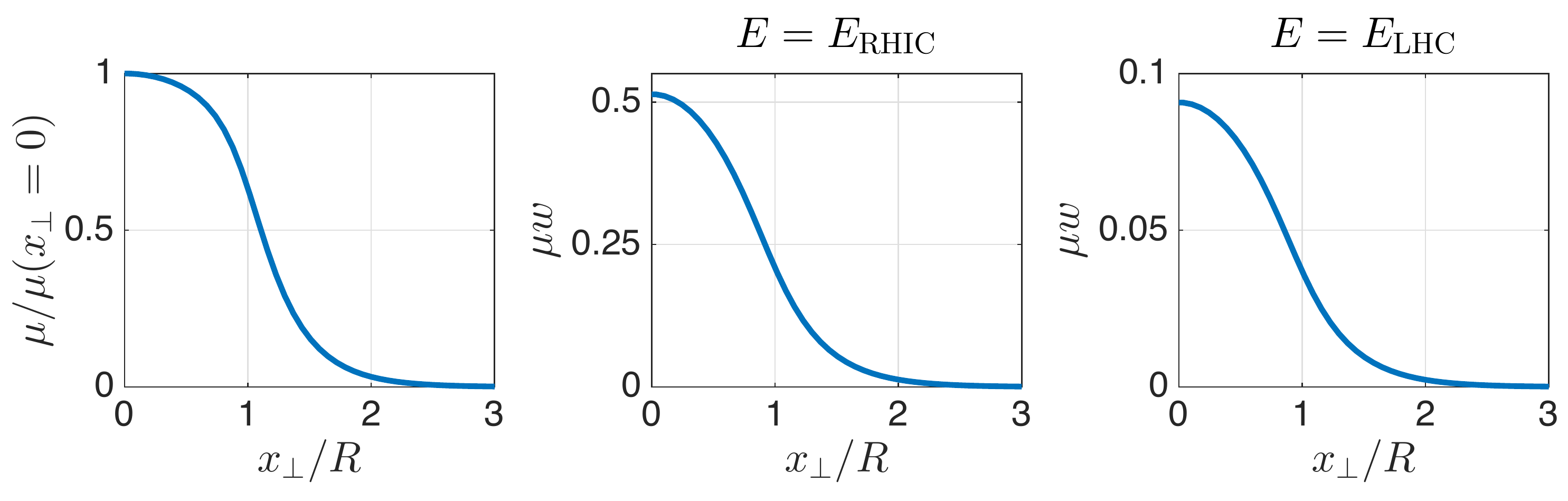}
\end{center}
\vskip -0.1in
\caption{%
Left: the normalized scale factor $\mu$
as a function of the transverse coordinate.  Note that 
the scale over which $\mu$ varies is the nuclear radius $R$.
Right: the dimensionless local shock thickness $\mu w$ as a function of 
transverse coordinate for both RHIC and LHC simulations.  Note that for our simulations $\mu w < 0.51$,
which is in the range of universality for the hydrodynamic initial data seen in Figs.~\ref{fig:initialhydrodata} and \ref{fig:rescaledproperenergy},
where $\mu w \leq 0.525$.
\label{fig:muandwidths}
} 
\end{figure}

We choose 
nuclear radius $R = 6.5$ fm and surface thickness $a = 0.66$ fm.  
We employ two different energies: 
$E = E_{\rm RHIC} = 200 \, {\rm GeV} \times \frac{N_{\rm Au}}{2}$
and
$E = E_{\rm LHC} = 2.76\,{\rm TeV} \times \frac{N_{\rm Pb}}{2}$ 
where $N_{\rm Au} = 197$ and $N_{\rm Pb} = 207$ are the number of nucleons in gold and lead nuclei respectively. 
These energies are energies of heavy ion collisions at RHIC and the LHC, where gold and lead nuclei are collided, 
and as such we simply refer to the resulting hydrodynamic simulations as ``RHIC" and ``LHC."   
We set the number of colors $N_{\rm c} = 3$
and $\gamma = \gamma_{\rm RHIC} = \frac{E_{\rm RHIC}}{m_{\rm N} N_{\rm Au}} \approx 100$ for the RHIC simulation 
and $\gamma = \gamma_{\rm LHC} = \frac{E_{\rm LHC}}{m_{\rm N} N_{\rm Pb}} \approx 1400$ for the LHC simulation. 
Here $m_{\rm N}\approx 1 $ GeV is the nucleon mass.  

Before continuing let us first ask whether the parameters in the previous paragraph 
yield collisions with small transverse gradients and with suitably small longitudinal widths 
as to enjoy the universal features of the hydrodynamic flow discovered in this paper.
First consider the size of transverse gradients of $\mu(\bm x_\perp)$.
Fig.~\ref{fig:muandwidths} shows a plot of $\mu/\mu(\bm x_\perp = 0)$ 
as a function of transverse coordinate.  Clearly the scale $\ell$ over which $\mu$ varies is $\ell \sim R$.
For comparison, for the RHIC simulation $\mu(\bm x_\perp = 0) R \approx 86$ and 
for the LHC simulation $\mu(\bm x_\perp = 0) R \approx 209$.  We therefore see that 
the separation of scales (\ref{eq:sepofscales}) is satisfied for both sets of collisions,
which justifies the use of the transverse gradient expansion.
Turning now to the local shock thickness, also included in Fig.~\ref{fig:muandwidths} 
are plots of $\mu(\bm x_\perp) w(\bm x_\perp)$ for both energies $E_{\rm RHIC}$
and $E_{\rm LHC}$.  We see that $\mu w$ takes its maximum value $0.51$
for $E = E_{\rm RHIC}$.  Therefore, the local widths are in the range of universality 
for the hydrodynamic initial data seen in Figs.~\ref{fig:initialhydrodata} and \ref{fig:rescaledproperenergy}, where $\mu w \leq 0.525$.
This justifies using the universal planar shock stress $\mathcal T^{\mu\nu}_{\rm planar}$ to 
construct initial hydrodynamic data in (\ref{eq:hydroIC}).

We construct our initial hydrodynamic data at time $\tau_{\rm init} = 5/\mu(\bm x_\perp = 0)=0.045 R \approx 0.3$ fm/c for the RHIC simulation 
and $\tau_{\rm init} = 7.5/\mu(\bm x_\perp = 0)=0.028 R \approx 0.2$ fm/c for the LHC simulation.  We then evolve forward in time using 
Israel-Stewart hydrodynamics \cite{Luzum:2008cw} with viscosity (\ref{eq:viscosity}) and relaxation time $\tau_{\Pi} = (2 - \log(2))/(2\pi T)$
 \cite{Bhattacharyya:2008jc,Baier:2007ix}.  For simplicity we focus on rapidity dependent observables only
 and leave a detailed analysis for future work.  
 
 \begin{figure}[h]
\vskip 0.15in
\begin{center}
\includegraphics[scale = 0.35]{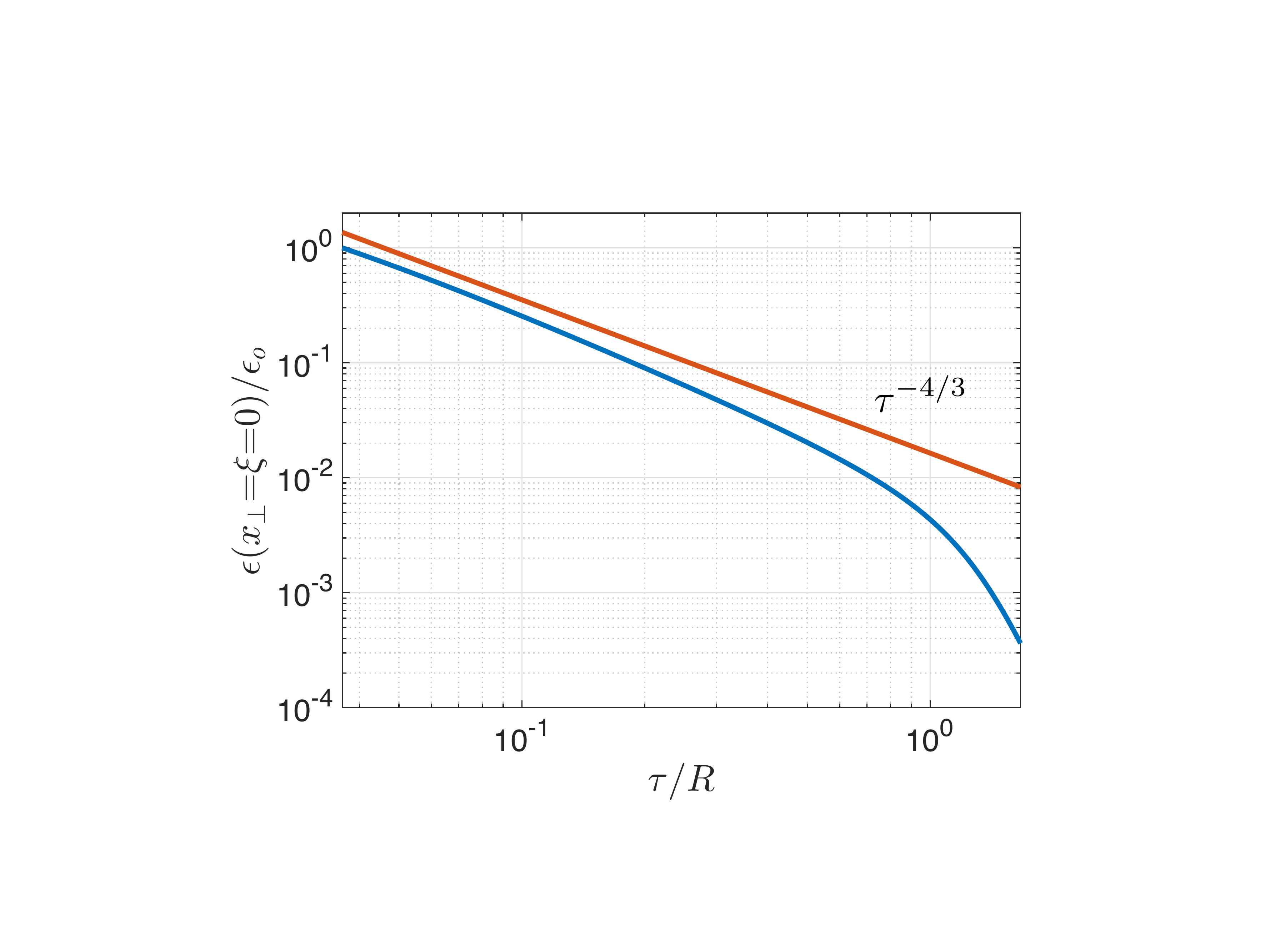}
\end{center}
\vskip -0.1in
\caption{%
The proper energy $\epsilon$ at $x_\perp = \xi = 0$ as a function of proper time. 
The normalization $\epsilon_o \equiv \epsilon(\tau = \tau_{\rm init},x_\perp = \xi = 0)$.
Initially the proper energy falls off like $\tau^{-4/3}$, just like boost invariant flow.
However, as time progresses the rate of fall off increases.  
\label{fig:CompareToBIF}
} 
\end{figure}

\begin{figure}[h]
\vskip 0.15in
\begin{center}
\includegraphics[scale = 0.45]{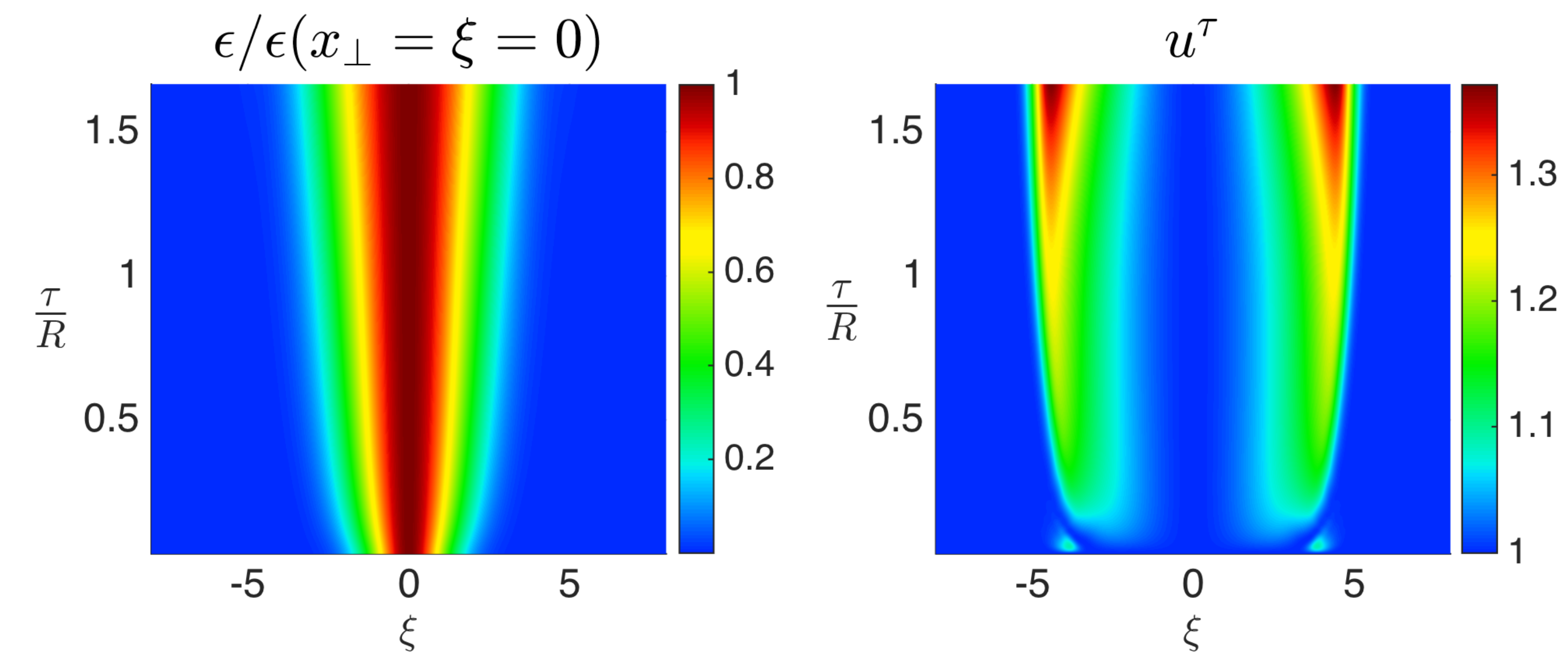}
\end{center}
\vskip -0.1in
\caption{%
Left: the normalized proper energy $\epsilon/\epsilon(\bm x_\perp = \xi = 0)$ at $\bm x_\perp = 0$ as a function of rapidity and proper time.
Right: the $\tau$ component of the fluid velocity at $\bm x_\perp = 0$ as a function of rapidity and proper time.
\label{fig:hydroflow}
} 
\end{figure}

Fig.~\ref{fig:CompareToBIF} shows a plot of the proper energy at $\bm x_\perp = \xi = 0$ as a function of $\tau$ for the LHC simulation.
Also included in the plot is the curve $\tau^{-4/3}$.  Note that for boost invariant flow 
the proper energy decays like $\tau^{-4/3}$.
At early times we see from the figure that $\epsilon \sim \tau^{-4/3}$.
However, as time progresses the rate of fall off grows faster than $\tau^{-4/3}$.  
Fig.~\ref{fig:hydroflow} shows plots of $\epsilon$ and $u^\tau$ at $\bm x_\perp = 0$ for the LHC simulation.
As time progresses the rapidity width of $\epsilon$ broadens.  Both the broadening and violation of the 
$\tau^{-4/3}$ scaling are due to the fact that $\epsilon$ has non-trivial rapidity dependence.  
Simply put, rapidity gradients drive longitudinal expansion 
faster than boost invariant flow, which results in the broadening of $\epsilon$ in $\xi$ and correspondingly, less energy at smaller rapidities
than there would be in the case of boost invariant flow.  Moreover, by time $\tau \sim R$, transverse gradients 
result in significant transverse expansion, which further enhances the violation of boost invariant flow.
The late-time violation of boost invariant flow also manifests itself in the fluid velocity.
At early times $u^\tau \approx 1$ and the fluid velocity is approximately that 
of boost invariant flow.
However, as time progresses deviations from $u^\tau = 1$ grow both in amplitude and domain.

\section{Spectrum of produced particles}
\label{sec:spectrum}

After the quark-gluon plasma produced in heavy ion collisions cools below the 
QCD deconfinement transition the system transitions from a quark-gluon liquid into a gas of hadrons.
An interesting observable to study is the spectrum of produced hadrons.
Using a Cooper-Frye freeze-out prescription, the spectrum of produced 
particles can be computed from the hydrodynamics evolution.  The spectrum of hadrons of
degeneracy $d$, four-momentum $p^\mu = (\mathcal E,\bm p)$
is given in terms of the hydrodynamic variables $\epsilon$ and $u^\mu$ by \cite{Cooper:1974mv}
\begin{equation}
\label{eq:CF1}
\mathcal E \frac{dN}{d^3 p} =  \frac{d}{(2 \pi)^3} \int d\Sigma^\mu p_\mu f(u^\mu p_\mu),
\end{equation}
where 
\begin{equation}
f(u \cdot p) = \frac{1}{\exp \left ({\frac{u \cdot p}{T_{\rm freeze}}} \right ) \pm 1},
\end{equation}
with the $+$ sign for Fermions and the $-$ sign for Bosons.  The integration in 
(\ref{eq:CF1}) is over the hypersurface of constant temperature $T = T_{\rm freeze}\approx 150$  MeV
with $T$ given in terms of the proper energy by (\ref{eq:Tdef}).  
In this simple study we assume all particles are massless bosons.  As such, the number of particles produced per unit rapidity is given by 
\begin{equation}
\frac{dN}{dy} = \frac{d}{(2 \pi)^3} \int d^2 p_T \int d\Sigma^\mu p_\mu f(u^\mu p_\mu),
\end{equation}
where the transverse component of the particle's momentum is $\bm p_T$ and $y = \tanh^{-1} \frac{p_z}{\mathcal E}$ is its rapidity
(which equals its pseudo-rapidity, as we are assuming massless particles).

\begin{figure}[h]
\vskip 0.15in
\begin{center}
\includegraphics[scale = 0.45]{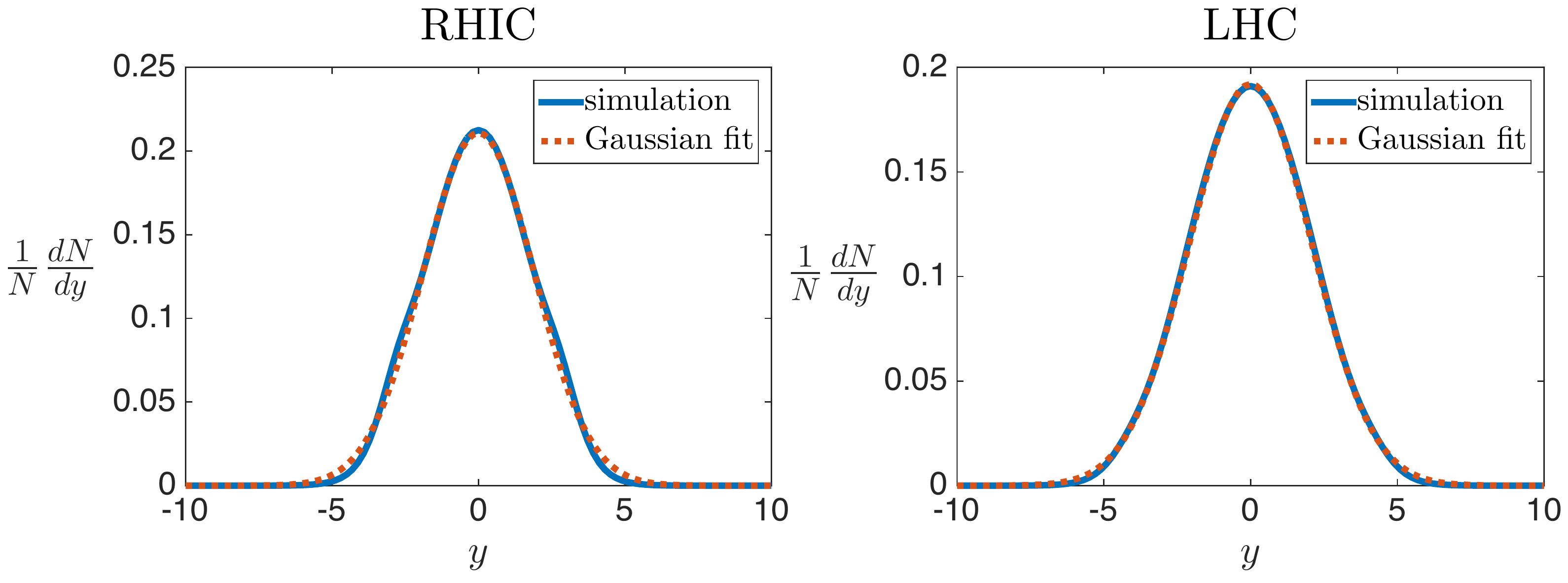}
\end{center}
\vskip -0.1in
\caption{%
The normalized rapidity distribution of particles for RHIC (left) and LHC (right) simulations together with Gaussian fits.  Both distributions are 
well approximated by Gaussians with width $\sigma = 1.9$ for the RHIC simulation and $\sigma = 2.1$ for the LHC simulation.
\label{fig:particlespectrumwithgaussian}
} 
\end{figure}

Fig~\ref{fig:particlespectrumwithgaussian} shows a plot of the normalized 
rapidity distribution of particles for our RHIC and LHC simulations together with Gaussian fits.
Both distributions are well described by Gaussians with rapidity widths $\sigma = 1.9$ 
for the RHIC simulation and $\sigma = 2.1$ for the LHC simulation.  
Curiously, the rapidity spectrum of particles produced in 200 GeV collisions at RHIC  
is also well approximated by a Gaussian with a width just 15\% larger than we observe in our holographic simulations 
\cite{Bearden:2004yx}.

\section{Concluding remarks}
\label{sec:conclusions}

Our results demonstrate that the hydrodynamic flow produced in strongly coupled collisions 
is insensitive to the detailed structure of the colliding shocks and has universal characteristics.
Finite $w$ effects, which can be appreciable in size, merely alter the normalization and 
rapidity width of the proper energy as opposed to changing its functional form altogether.
One utility of this observation is that numerical simulations of collisions need not have 
asymptotically small shock widths in order to approach the $\delta$-function limit.  This observation is especially valuable 
for simulations without any spacetime symmetry, such as the off-center shock collisions of \cite{Chesler:2015wra},
where taking the shock thickness $w \to 0$ can be computationally expensive.  

We note, however, that when the width $w$ becomes of order the microscopic relaxation time in the produced plasma, which is
$t_{\rm hydro}\approx 2/\mu$,
the structure of the shock can imprint itself on the hydrodynamic evolution.   
Indeed, as seen in Fig.~\ref{fig:universalhydro}, when $\mu w \sim 1/2$ the proper energy begins 
to develop small deviations from the universal behavior in Eq.~(\ref{eq:gaussianform0}).  
Why does this happen?  When $w \gtrsim \mu$, the system cannot equilibrate 
until energy stops piling up in the collision plane, which happens for an amount of time of order $w$. 
The resulting hydrodynamic evolution must then become sensitive to the detailed structure of the shocks.
Indeed, it was demonstrated in \cite{Casalderrey-Solana:2013aba} that for sufficiently wide shock collision, the 
dynamics are well described by the Landau model of heavy ion collisions, 
where the nuclei are assumed to be thermalized at the time they overlap completely and the initial fluid velocity is small \cite{Landau:1953gs,Steinberg:2004vy,Wong:2008ex}.

The observation that the proper energy rapidity width 
has a finite limit as the shock width $w$ vanishes implies 
that at $w = 0$, the only source of rapidity broadening comes from hydrodynamic evolution 
alone, where rapidity gradients drive longitudinal expansion.  This stands in contrast to 
asymptotically free QCD, where the longitudinal thickness of nuclei deceases due to Lorentz contraction
and the rapidity width of the produced plasma grows larger and larger as the energy is increased \cite{Gelis:2010nm}.   It would be interesting to study 
finite coupling corrections to our strongly coupled collisions and see how they affect 
the initial hydrodynamic data.

\acknowledgments

We thank Jorge Casalderrey-Solana, Michal Heller, Krishna Rajagopal, Paul Romatschke, Bj\"{o}rn Schenke and Larry Yaffe for useful discussions.
PC is supported
by the Fundamental Laws Initiative of the Center for the
Fundamental Laws of Nature at Harvard University.  
NK is supported by the German National Academic Foundation and the Elite Network Bavaria.
WS is supported by the U.S. Department of Energy under grant Contract
Number DE-SC0011090.



\end		{document}